\documentclass{article}[12pt]

\usepackage[utf8]{inputenc}
\usepackage[T1]{fontenc}
\usepackage{xcolor}
\usepackage{graphicx}
\usepackage{hyperref}
\usepackage{subcaption}
\usepackage{tikz}
\usepackage{slashbox}

\hypersetup{%
colorlinks=true,
breaklinks=true,
linkcolor=red,
anchorcolor=black,
citecolor=brown,
filecolor=blue,
menucolor=red,
urlcolor=red,
}
\usepackage[round,longnamesfirst]{natbib}
\usepackage{mathtools, amsfonts, amssymb,amsthm,stmaryrd,amsmath}
\mathtoolsset{showonlyrefs}
\usepackage{enumerate}
\usepackage{comment}
\usepackage{ifthen}
\usepackage{mathrsfs}
\usepackage{bbm}
\usepackage[symbol*]{footmisc}
\setfnsymbol{wiley}
\usepackage{booktabs}
\usepackage{algorithm}
\usepackage{algpseudocode}

\usepackage{xr}
\newcounter{prop}
\newtheorem{proposition}[prop]{Proposition}

\newcounter{def}
\newtheorem{definition}[def]{Definition}
\newcounter{rem}
\newtheorem{remark}[rem]{Remark}

\title{Rate accelerated inference for integrals of multivariate random functions}
\author{Valentin Patilea\footnote{Ensai, CREST - UMR 9194, France; valentin.patilea@ensai.fr} \qquad Sunny G.W. Wang\footnote{Ensai, CREST - UMR 9194, France; sunny.wang@ensai.fr}
}
\date{\today}

\begin{document}

\maketitle

\begin{abstract}
The computation of integrals is a fundamental task in the analysis of functional data, which are typically considered as random elements in a space of squared integrable functions. Borrowing ideas from recent advances in the Monte Carlo integration literature, we propose effective unbiased estimation and inference procedures for integrals of uni- and multivariate random functions. Several applications to key problems in functional data analysis involving random design points are studied and illustrated. In the absence of noise, the proposed estimates converge faster than the sample mean and the usual algorithms for numerical integration. Moreover, the proposed estimator facilitates effective inference by generally providing better coverage with shorter confidence and prediction intervals, in both noisy and noiseless setups.

\end{abstract}

\quad 


\qquad

\section{Introduction}

Functional data analysis (FDA) is an increasingly important field of statistics that
supplies useful methodology for the analysis of data whose datum are functions.
The complexities of data sets have grown in tandem with the increasing
sophistication of data collection mechanisms. An increasing number of
applications feature functional data collected at a discrete, random set of design
points, also known as the random design framework. Examples include sports
science \cite{leroy2023}, \cite{warmenhoven24}, oceanography \cite{acar18},
\cite{yarger22}, medicine  \cite{sorensen2013}, and spatial data
\cite{Burbano2024}. In these applications, a fundamental task in the functional
data analysis pipeline is the approximation of integrals of functions that depend on the sample paths (also called trajectories).

Despite their importance, the approximation of integrals are often treated with
secondary importance, with practitioners often resorting to simple methods such
as Riemann sums or sample means. This is often sub-optimal in terms of accuracy
and inappropriate for the purposes of inference, when the goal is to construct
prediction or confidence intervals. On the one hand, when the sample paths are
observed without noise, faster rates of convergence can be attained. On the other
hand, when the observations are contaminated with noise, using an integral
approximation method with slower or comparable rate to $M^{-1/2}$ (where $M$ is
the number of design points) affects the asymptotic variance. This complicates the
construction of confidence intervals, since the asymptotic variance depends on the
trajectories.

In this paper, we propose an integral approximation approach over compact domains, specifically
designed to address the demands of statistical inference in FDA. It satisfies
several important criteria: (i) achieve faster rates of convergence relative to
existing methods; (ii) enable the simple and effective construction of short
prediction and confidence intervals with desired coverage levels; (iii) be flexible
enough to accommodate multivariate domains, such as cubes or spheres, of the design points; (iv) be
computationally fast; (v) encompass both noisy and noiseless situations with
minimal adjustment required by practitioners; and (vi) be adaptive to the regularity
of the functions’ data generating process. The last property is a by-product of
recent work in regularity estimation; see for example \cite{golo1}, \cite{WKP},
\cite{kassi2023}.

Our contribution builds upon recent advances in the field of Monte Carlo
integration, tailoring its methodology for the field of FDA. Our proposed
methodology is general, and applies to many different contexts in FDA, such as
functional regression, scores and data depths. Equipped with our \textbf{R}\footnote{Available at https://github.com/sunnywang93/integratefda}
package, minimal effort is required from the practitioner to adapt to a wide array of
data and modeling settings in FDA.

The paper is organized as follows. Section \ref{sec:motivation} formally motivates the integration problem in FDA, discussing the most commonly used approaches and their limitations. Section \ref{sec:methodo} describes the proposed estimation and inference procedures, elucidated with several concrete examples in Section \ref{sec:appli}. Section \ref{disc_beta} discusses the issue of random integrands unique to the context of FDA, and the subtle concerns related to the regularity of the sample paths. Section \ref{sec:numeric} describes an extensive simulation study exploring the finite sample properties of the proposed methodology. Finally, we apply our methodology to analyze swimmers' performance curves in Section \ref{sec:data-appli}.

 

\section{Motivation and problem formulation}\label{sec:motivation}
Let $X = \{X(t) : t \in \mathcal{T}\}$ be a second order stochastic process defined on a compact
domain $\mathcal{T}$. The typical examples we have in mind are $\mathcal T = [0,1]^d$, the unit  cube in $\mathbb R^d$, $d\geq 1$, and $\mathcal T = \mathcal S^d$ the Euclidean unit sphere in $\mathbb R^{d+1}$. The methodology presented below accommodates for vector-valued sample paths, that is $X(t) \in\mathbb R^K$, $\forall t\in\mathcal T$, for some $K\geq 1$. However, for simplicity, if not stated differently, we consider $K=1$.

In this paper, we will focus on the so-called random design framework arising in many applications, where the sample paths of $X$ are observed at random and discrete points, possibly contaminated with noise. These observations come in the form of pairs $(Z_{i,m}, T_{i,m}) \in \mathbb{R} \times \mathcal{T}, 1 \leq i \leq n, 1 \leq m \leq M_i$, where $M_i$ is a possibly random, positive integer. We refer to $T_{i,1},\dots, T_{i,M_i}$ as the design points, which are independent copies of a random variable $T$. The observed pairs are generated under the model
\begin{equation}\label{ini_mod}
Z_{i,m} = X_i(T_{i,m}) + \sigma(T_{i,m}; X_i(T_{i,m}))e_{i,m}, \qquad 1 \leq m \leq M_i, 1 \leq i \leq n,
\end{equation}
where $X_i  $  are independent sample paths of $X$, and the error terms $e_{i,m}$ are independent copies of a random variable $e$ with zero mean and unit variance. We assume that $X$, $T$ and the $M_i$'s are mutually independent. The noiseless (resp. noisy) case corresponds to a null (resp. positive) conditional variance function $\sigma^2(\cdot;\cdot)$. Both cases will be studied in the following sections.

The usual paradigm is to consider $X$ as a random function taking values in the space of squared integrable functions endowed with the $L^2(\mathcal T)-$inner product. Many problems in FDA then involve computing an integral of a functional  over the domain $\mathcal T$ with the functional depending on the sample of $X$, which is then taken as given. To formalize the integral calculation problem, let $\rho$ be the probability distribution of $T\in \mathcal{T}$. Given a sample path, consider the integral functional of the form
\begin{equation}\label{int-problem}
	I(\varphi) = \int_{\mathcal{T}} \varphi  \ d\rho =: \mathbb{E}\left[\varphi \right],
\end{equation}
where $\varphi:\mathcal{T} \rightarrow \mathbb{R}$ is a function of the sample path, and $\varphi(t)$ is a short notation for $\varphi(t, X(t))$. Thus, here $\mathbb{E}\left[\varphi \right]$  is a simple notation for  $\mathbb{E}\left[\varphi (T,X(T))\mid X\right]$. Examples include performing out-of-sample prediction, estimating fPCA scores, and computing data depths. More details can be found in Section \ref{sec:appli}.  In order to simplify the following exposition, where there is no danger of confusion, we will simply denote a sample path by $X$, and the associated design points by $T_1,\ldots,T_M$.

A common approach in practice to approximate \eqref{int-problem} when $\mathcal{T}$ is a compact interval on the real line,  is to use  Riemann sums, such as the trapezoidal rule. If $T_{(m)}$ is the $m$-th order statistic of $T_1,\ldots,T_M$, the trapezoidal rule is given by 
\begin{equation}
\widehat I^{\rm trapez}(\varphi) = \sum_{m=1}^{M+1} \frac{\varphi(T_{(m-1)}) + \varphi(T_{(m)})}{2} \left\{T_{(m)} - T_{(m-1)}\right\},
\end{equation}
with some rule for the endpoints, for example $T_{(0)}= \min \mathcal T$, $\varphi(T_{(0)})=\varphi(T_{(1)})$, and $T_{(M+1)}= \max \mathcal T$, $\varphi(T_{(M+1)})=\varphi(T_{(M)})$. If $\varphi$ is $\beta$-Hölder, $\widehat I^{\rm trapez}(\varphi)$ attains the convergence rate $O_{\mathbb{P}}(M^{-\beta}), \beta \in (0, 1]$. 
In the multivariate case, that is, if $\mathcal{T} \subset \mathbb{R}^d$ with $d>1$, the construction of random Riemann sums is in practice non-trivial and requires a careful partitioning of the domain \citep[see, e.g.,][]{pruss1996randomly}. Their expected rate of convergence is $O_{\mathbb{P}}(M^{-\beta/d})$. 

With the random design points $T_1,\ldots,T_M$, an alternative approach, sometimes called Monte Carlo integration, is to estimate the integrals using a sample mean. This gives estimates with a convergence rate of $O_{\mathbb{P}}(M^{-1/2})$, independent of the dimension. Although the Central Limit Theorem guarantees convergence in the distribution, inference based on the asymptotic distribution may be inaccurate if $M$ is not sufficiently large.

Using recent developments in Monte Carlo integration, estimates of $I(\varphi)$ with the rate $O_{\mathbb{P}}(M^{-1/2-\beta/d})$ can be obtained.
Such faster rates are obtained under the same Hölder continuity assumptions used for the random Riemann sums. The approach proposed below improves the estimation of key quantities in FDA and allows for effective inference. In the absence of noise, very short prediction intervals can be constructed. In the case of noisy observations, the integration error is negligible with respect to the convergence of the distribution, allowing a simple construction of confidence intervals.

\section{Methodology}\label{sec:methodo}

Our methodology for constructing estimates of  $I(\varphi)$ is based on the so-called control variates approach. See \cite{oates2017control}, \cite{novak2016some}, \cite{BAKHVALOV2015502}. We first recall the general principle of the control variates approach, and next, the one based on nearest neighbors elaborated by \cite{leluc2024speeding}. Finally, we construct inference for $I(\varphi)$ in both noiseless and noisy cases. 

\subsection{Control variates, the principle}\label{sec:control-variate}
To briefly describe the key elements, we start from the general principle of using control variates. Let $\varphi: \mathcal{T} \rightarrow \mathbb{R}$ be a given, generic $\beta$-Hölder integrand as in  \eqref{int-problem}, with observed values $\varphi(T_m), 1 \leq m \leq M$. The $T_m$ are random, and $M$ can be random too. The key idea of control variates is to reduce the variance of the sample mean estimate by centering the expectation using a suitable function whose integral is known. Let $\widetilde \varphi$ be a generic approximation of $\varphi$, called a control variate, whose integral $I(\widetilde \varphi)$ can be explicitly calculated. Let $\mathbb{E}_M[\cdot] = \mathbb{E}[\cdot \mid M]$ and $\operatorname{Var}_M[\cdot] = \operatorname{Var}[\cdot \mid M]$ denote the conditional mean and variance given $M$, respectively. The integral can be written as 
\begin{equation}
I(\varphi) = \mathbb{E}_M[\varphi] - \mathbb{E}_M\left\{\widetilde \varphi - \mathbb{E}_M[\widetilde \varphi]\right\}.
\end{equation}
A natural estimator is given by 
\begin{equation}\label{eq:control-variate-gen}
\widehat I(\varphi) = \frac{1}{M}\sum_{m=1}^M\big\{\varphi(T_m) - \left[\widetilde \varphi(T_m) - I(\widetilde \varphi ) \right]\big\}.
\end{equation}
Let $|\cdot|_{\infty}$ denote the uniform norm, and let $\lesssim$ mean the left side is bounded by a constant times the right side. By construction, we have
\begin{equation}
\mathbb{E}_M\left[\widehat I(\varphi)\right] = I(\varphi), \qquad \text{and} \qquad \operatorname{Var}_M\left[\widehat I(\varphi)\right] \lesssim M^{-1}|\varphi - \widetilde \varphi|^2_{\infty},
\end{equation}
so the estimate remains unbiased. Moreover, choosing a control variate $\widetilde \varphi$ that is sufficiently close to $\varphi$ in terms of the uniform norm, the variance is reduced, leading to faster rates of convergence. 

\subsection{Control variates with nearest neighbor}
Using the leave-one-out nearest neighbor as a control variate is proposed in \cite{leluc2024speeding}. See also \cite{oates2017control}. Let $d(\cdot,\cdot)$ be a distance on $\mathcal T$. In the following, for simplicity, we  consider either  $\mathcal T = [0,1]^d$, the unit cube in $\mathbb R^d$, or $\mathcal T = \mathcal S^d$, the unit sphere in $\mathbb R^{d+1}$. Then the distance $d(\cdot,\cdot)$ is either the Euclidean distance or the geodesic distance.

Let $m \in \{1, \dots, M\}$,  $\mathbf{T} =  \mathbf{T}_M = (T_1, \dots, T_M)$, $\mathbf{T}^{(m)} = \mathbf{T}^{(m)}_M = \mathbf{T} \setminus \{T_m\}$. The leave-one-out nearest neighbor (LOO-NN) is given by 
\begin{equation}\label{NN_t}
\widehat N^{(m)}(t) = \widehat N_M^{(m)}(t) \in \arg\min_{s \in \mathbf{T}^{(m)}} d(t, s),
\end{equation}
where any ties are broken with lexicographic order. The unbiased estimator in \eqref{eq:control-variate-gen} is then given by 
\begin{equation}\label{control-neighbors-estim}
\widehat I(\varphi) = \frac{1}{M}\sum_{m=1}^M \left\{\varphi(T_m) - \left[\widetilde \varphi^{(m)}(T_m) - I\left(\widetilde \varphi^{(m)}\right) \right] \right\},
\end{equation}
where $\widetilde \varphi^{(m)}(t) = \varphi(\widehat N^{(m)}(t))$ denotes the function $\varphi$ evaluated at its leave-one-out nearest neighbor.

\medskip

\begin{proposition}\label{th1_leluc}\cite[Theorem 1,][]{leluc2024speeding} 
Assume that $M\geq 4$ and   $T_1,\ldots,T_M$  are random copies of $T\in\mathcal T $ which admits a density $f_{  T}$ for which constants  $C_0, C_1 $ exist such that $0< C_0 \leq  f_{ T} \leq C_1$. Moreover, $\varphi$ is $\beta-$Hölder, that is constants $L_\varphi >0$ and $\beta \in (0,1]$ exist such that 
$$
|\varphi (t) - \varphi(s) |\leq L_\varphi d(t,s),\qquad \forall s,t\in\mathcal T.
$$
Then a constant $C_{\rm NN-loo}$ exists, depending only on $\beta$, $L_\varphi$, $C_0$, $C_1$ and $d$, such that, for  $\widehat I(\varphi)$ in \eqref{control-neighbors-estim},
\begin{equation}\label{E_M}
\operatorname{Var}_M\left[\widehat I(\varphi)\right]^{1/2} = \mathbb{E}_M\left[\left|\widehat I(\varphi) - I(\varphi) \right|^2 \right]^{1/2} \leq C_{\rm NN-loo} M^{-1/2}M^{-\beta/d}.
\end{equation}
\end{proposition}

\medskip

The rate in Proposition \ref{th1_leluc} is known to be the optimal rate, see \cite{novak2016some}, \cite{BAKHVALOV2015502}. The fastest possible rate is obtained for  Lipschitz functions, \emph{i.e.,} $\beta = 1$, when  the bound in \eqref{E_M} becomes  $M^{-1/2-1/d}$. It is worth noting that the rate in in Proposition \ref{th1_leluc} also  holds for sets $\mathcal T$ in more general metric spaces, as proved by \cite{leluc2024speeding}.

The control neighbors unbiased estimator has the attractive feature of being a linear integration rule
\begin{equation}\label{weighted-cn}
\widehat I(\varphi) = \widehat I(\varphi(\mathbf{T})) = \sum_{m=1}^M w_{M,m} \varphi(T_m),
\end{equation}
where the explicit form of the weights, depending only on the $T_m$'s, is given in \cite{leluc2024speeding} and the Appendix for completeness. The expressions \eqref{control-neighbors-estim} and \eqref{weighted-cn} are used interchangeably throughout the paper.

\subsection{Inference of integral estimates}\label{sec:noiseless-inf}
We use the control variates idea  with leave-one-out nearest neighbor for approximating functionals in the context of FDA. This leads to approximation with faster rates compared to the common approaches by random Riemann sums or sample means. For the inference, we have to distinguish between the noiseless ($\sigma^2 =0$) and noisy ($\sigma^2 >0$) cases. In the latter case, the noise is expected to drive the inference for $I(\varphi)$ because the integral approximation has a faster rate of decrease than that given by the Central Limit Theorem (CLT).

Let us first consider the case where the sample paths are observed without noise. The estimator in \eqref{control-neighbors-estim} was shown to converge at a rate of $M^{-1/2-\beta/d}$ in probability, however its convergence in distribution remains an open question. We therefore propose prediction intervals for $\widehat I(\varphi)$ using a $M^*$-out-of-$M$ subsampling procedure. Assume for the moment that the rule for $M^*$ and the regularity parameter $\beta$ are given.  Let $\mathbf{T} =  \mathbf{T}_M = (T_1, \dots, T_M)$ denote the vector of random sampling points, and $\varphi(\mathbf{T}) = (\varphi(T_1), \dots, \varphi(T_M))$ be the vector of observed values of the functional. Let $B$ be some large integer and $1 - \delta$ be the coverage level, both chosen by the practitioner. In our simulation experiences we take $B$ to be 1000.  Given the pairs $(\mathbf{T}, \varphi(\mathbf{T}))$, and the unbiased estimate $\widehat I(\varphi(\mathbf{T})) $ computed according to \eqref {weighted-cn}, Algorithm \ref{algo:pred-intervals} can be used to construct prediction intervals centered at $\widehat I(\varphi(\mathbf{T})) $.

\medskip

\begin{algorithm}
\caption{Prediction Intervals for Control Neighbor Estimates}
\label{algo:pred-intervals}
\begin{algorithmic}[1]
\Require Data $(\mathbf{T}, \varphi(\mathbf{T}))$, Integral estimate $\widehat I(\varphi(\mathbf{T}))$, Replications $B$, Confidence level $1-\delta$, Subsample size $M^*$, Regularity  $\beta$

Initialize $I_{B, M^*}  \gets \emptyset$;

\For{{$b = 1, \dots, B$}}

\State $\mathbf{T}^*_{B, M^*} \gets (T^*_{B, 1}, \dots, T^*_{B, M^*})$;  \Comment Sample $M^* < M$ points from $\mathbf{T}$ without replacement;

\State Compute $\widehat I(\varphi(\mathbf{T}^*_{B, M^*}))$ using \eqref{control-neighbors-estim}; \Comment Integral estimate with the subsample;

\State $I_{B, M^*} \gets I_{B, M^*} \bigcup \widehat I(\varphi(\mathbf{T}^*_{B, M^*}))$; \Comment Store  integral estimate by subsampling;

\EndFor

\State Compute $q_{\delta/2}$ and $q_{1 - \delta/2}$ empirical quantiles of $(M^*)^{1/2 + \beta/d}\left[\widehat I(\varphi(\mathbf{T}^*_{B, M^*})) - \widehat I(\varphi(\mathbf{T})) \right]$;

\State Set
$$\text{PI}_{1 - \delta} := \left[\widehat I(\varphi(\mathbf{T})) + M^{-1/2 - \beta/d}q_{\delta/2}, \;\;\widehat I(\varphi(\mathbf{T})) + M^{-1/2 - \beta/d}q_{1 - \delta/2}  \right];$$

\State \Return $\text{PI}_{1-\delta}$;

\end{algorithmic}
\end{algorithm}

\medskip

We conjecture that the prediction interval $\text{PI}_{1 - \delta} $ has the asymptotic level $1 - \delta$ under the conditions of Proposition \ref{th1_leluc} and with a suitable rule for $M^*$, as $M$ increases.

The data-driven parameters to be chosen are the Hölder exponent $\beta$ and the subsample size $M^*$. We argue that a reasonable choice for $M^*$ is to set $M^* = \lfloor M/2 \rfloor$. This choice allows for the largest number of distinct subsamples, which according to Stirling's formula is about $2^{2M}/\sqrt{M\pi}$ when $M$ is large. The idea of half-sample subsampling can also be found in a similar context in the literature on bagging, see \cite{buja2006}, \cite{buhlmann2022}. 

\begin{remark}
	It is worth noting that the length of the prediction intervals generated by the Algorithm \ref{algo:pred-intervals} is $O_{\mathbb P} (M^{-1/2}) \times O_{\mathbb P} (M^{-\beta/d})$, which is negligible compared to any common  method. Indeed, 
	the sample mean converges at the rate given by the Central Limit Theorem, \emph{i.e.,}  $O_{\mathbb P} (M^{-1/2})$, while the Riemann sums for $\beta-$Hölder functions defined on $d-$dimensional domains generally have the rate $ O_{\mathbb P} (M^{-\beta/d})$. 
\end{remark}

\begin{remark}
The choice of $\beta$ is more delicate, as it determines the rate of the length of the prediction interval. In asymptotic theory, the influence of smoothness is limited by the fact that $\beta \leq 1$. The simulations show that having a smoother integrand  $\varphi$ than just Lipschitz continuous still has some influence for moderate sample sizes $M$. 
\end{remark}
\begin{remark}\label{rem_reg}
It is worth recalling that in the common FDA applications,  $\varphi(t)$ stands for  $\varphi(t, X(t))$. More precisely, $\varphi$ is  a functional, usually smooth, of the sample path. Then the value of $\beta$ is given by the regularity of the sample paths of  $X$. In FDA, this regularity is often chosen in an \emph{ad-hoc} manner by examining, or simply imposing, the decay rate of eigenvalues. A better alternative is to estimate the Hölder exponent of the sample path of $X$. Recent contributions allow such an adaptive approach where $\beta$ is no longer imposed, but chosen in a data-driven way using a functional data sample. See \cite{golo1}, \cite{kassi2023}, \cite{WKP}. See also the discussion in Section \ref{disc_beta}. 
\end{remark}

\subsection{Inference with noisy integrands}\label{sec:inf-noisy-int}

In some applications, the measurements are contaminated with noise. Instead of observing the pairs $(T_m, \varphi(T_m)), 1 \leq m \leq M$ directly, one has access to noisy counterparts
\begin{equation}\label{noisy_int}
\phi(T_m) = \varphi(T_m) + \sigma_{\eta}(T_m) \eta_m, \qquad 1 \leq m \leq M, 
\end{equation}
where $\sigma_\eta (\cdot) \geq 0$, and $\eta_{m}$ are random copies of $\eta$, independent of the design points, and $\mathbb{E}[\eta ] = 0$ and  $\mathbb{E}[\eta^2   ] = 1$.
With noisy values $\phi(T_m)$ as in \eqref{noisy_int}, the feasible version of the unbiased estimator of $I(\varphi)$
is then
\begin{equation}\label{eq:control-neighbors-noisy}
\widehat I (\phi) = \sum_{m=1}^M w_{M,m} \phi(T_m) = I(\varphi) + \widehat \Sigma + R,
\end{equation}
where 
\begin{equation}\label{def_Sig}
\widehat \Sigma = \sum_{m=1}^M w_{M,m} \sigma_{\eta}(T_m)\eta_m  \qquad  \text{ and } \qquad R = \widehat I(\varphi) - I(\varphi),
\end{equation}
with the ideal $\widehat I(\varphi)$ defined according to \eqref{weighted-cn}. In the presence of noise, the rate of $\widehat I (\phi)$  is driven by $\widehat \Sigma$, as shown in the next convergence in distribution result. 

\medskip

\begin{proposition}\label{CLT-prop}
Assume the conditions of Proposition \ref{th1_leluc} hold true, with $\mathcal T = [0,1]^d$, $d\geq 1$. Assume that  $\eta_{1},\ldots,\eta_M$ are random copies of a zero-mean variable $\eta$ with unit variance and independent of the design points. Moreover, the conditional variance in \eqref{noisy_int} is such that $ 0<\inf_{t\in\mathcal T}\sigma_\eta(t) \leq \sup_{t\in\mathcal T}\sigma_\eta(t) <\infty$. Then, for $\widehat I (\phi)$ defined in \eqref{eq:control-neighbors-noisy}, it holds that 
	\begin{equation}
		\frac{1}{s_M} \left(\widehat I (\phi) - I (\varphi) \right) \stackrel{d}{\longrightarrow} \mathcal{N}(0, 1) \qquad \text{with} \qquad s_M^2 =  \sum_{m=1}^M w^2_{M,m}\sigma^2_{\eta}(T_m).
	\end{equation}
In the case  $\mathcal T=[0,1]$, we have $s_M^2 =   (5/2)M^{-1} \mathbb{E}_M\left[\sigma_\eta^2(T) \right] \left\{1 + o_{\mathbb P}(1) \right\}$, provided the density $f_T$ is $\alpha_f-$Hölder continuous for some $\alpha_f>0$.  
\end{proposition}

\medskip


\medskip

As a direct consequence of Proposition \ref{CLT-prop}, an asymptotic $(1 - \delta)-$level confidence interval for $I(\varphi)$  is
\begin{equation}\label{CI_noisy}
\text{CI}_{1 - \delta} = \left[\widehat I(\phi) - z_{1 - \delta/2}  s_M, \widehat I(\phi) + z_{1 - \delta/2}  s_M \right],
\end{equation}
where $z_{\delta}$ denotes the $\delta$-quantile of the standard normal distribution. 

\medskip

\begin{remark}
	In the framework defined by \eqref{noisy_int} where the $\varphi(T_m)$ are not directly observed, and $\varphi$ is $\beta-$Hölder for some $\beta >0$, it is no longer necessary to know  the regularity $\beta$. Indeed, the asymptotic interval $\text{CI}_{1 - \delta} $ is based on the asymptotic Gaussian distribution of $\widehat I (\phi)$, and  does not depend on $\beta$.
\end{remark}

\medskip

\begin{remark}
Since they are characterized by different regimes, we use a different terminology for the inference with noisy and noiseless integrands.  When the integrand is observed without noise, we refer to the intervals as prediction intervals, and denote them by $\text{PI}_{1 - \delta}$, see  Algorithm \ref{algo:pred-intervals}. When the integrand is observed with noise, we refer to it as confidence intervals instead, denoted by $\text{CI}_{1 - \delta}$, see \eqref{CI_noisy}. This distinction is used in all the examples discussed in the following.  
\end{remark}


\section{Applications}\label{sec:appli}
In this section, we present concrete examples of well-known applications in FDA for our approach to computing integral functionals using the control neighbors. The examples relate to functional regression, functional principal component  analysis (fPCA) and functional depths. The integral functions we present below depend on some unknown quantities such as the slope and the intercept in functional regression, the variance of the measurement error of the functional data, \emph{etc}. In order to focus on the novelty and the advantages of our approach compared to the existing ones, we take such quantities as given. As with any other approach to computing integral functionals, in real data applications we have to use some estimates for the unknown quantities. For all approximate methods of calculating integral functionals, the effect of these estimates is expected to disappear when the functional data set is large.


\subsection{Prediction and inference in functional regression models}\label{sec:frm}

\subsubsection{Functional linear model}\label{sec:pred-flm-noiseless}
Let $X(t)\in\mathbb R^K$, $t \in\mathcal T$, for some $K\geq 1$. That means, the sample paths are $K-$dimensional functions defined on a multivariate domain $\mathcal T$. Let $\langle .,. \rangle$ to be the standard inner product on $L^2(\mathcal{T})^K$; see \cite{Happ} for the formal definition. The functional linear model is given by
\begin{equation}\label{flm-model}
Y = \alpha_0 + \langle \alpha, X \rangle + \epsilon,
\end{equation}
where $(X, Y) \in L^2(\mathcal{T})^K \times \mathbb{R}$ is a random couple defined on a probability space, and $\epsilon$ is a random noise such that $\mathbb{E}(\epsilon\mid X) = 0$ and $\mathbb{E}(\epsilon^2\mid X) = \sigma^2_{\epsilon}(X)$. In the random (sometimes called independent) design framework where the values of $X_i$ are observed \emph{without} error, the observations in the learning set are in the form  
$$
(Y_i,X_i(T_{i,1})^\top ,\ldots,X_i(T_{i,M_i})^\top)^\top \in \mathbb R\times \underbrace{\mathbb R^K\times\cdots\times \mathbb R^K}_{\text{$M_i$ times}}, \quad  1 \leq i \leq n.
$$
(Here, the vectors are column matrices, and for a matrix $A$, $A^\top$ denotes the transpose.) We are interested in out-of-sample prediction of the response $Y_{n+1}$ using $X_{n+1}(T_{{n+1},m})$, $1\leq m \leq M_{n+1}$.
The $T_{{n+1},m}$ are random copies of $T\in\mathcal T$, independent of $X_{n+1}$ and $M_{n+1}$, and $T$ admits a density $f_T$. 

A wide variety of methods are available to learn the scalar $\alpha_0$ and the vector-valued function $\alpha$; see for example \cite{caihall2006}, \cite{crambespline}, \cite{comtefunctionalreg},  \cite{yuancairkhs2010}, \cite{caiyuan2012reg}, \cite{Zhou2022}. Our goal is not to revisit the estimation of $\alpha_0$ and $\alpha$, but rather to improve the out-of-sample prediction through an accurate estimation of the integrals. We thus treat these two quantities as given.

Assuming $(Y_{n+1},X_{n+1})$ follows the model \eqref{flm-model} and is independent of the learning sample, the best predicted mean  value of $Y_{n+1}$ given the sample path $X_{n+1}$ can be written as 
\begin{equation}\label{infeasible-reg-pred}
\widetilde Y_{n+1} = \alpha_0 + \mathbb{E}\left[\frac{\alpha(T)^\top X_{n+1}(T)}{f_T(T)}  \mid X_{n+1}\right],
\end{equation}
which can be approximated by the control variate estimate in \eqref{control-neighbors-estim} by
\begin{equation}\label{pred-noiseless}
\widehat Y_{n+1} = \alpha_0 + \sum_{m=1}^{M_{n+1}} w_{M_{n+1},m} \varphi(T_{n+1, m}), \quad \text{ with } \varphi(T_{n+1,m}) =  \frac{\alpha(T_{n+1,m})^\top X_{n+1}(T_{n+1,m})}{f_T(T_{n+1,m})}.
\end{equation}
In practice, the density $f_T$ can be estimated using non-parametric methods by pooling all the design points, resulting in $ \sum_{i=1}^n M_i$ points, much more than $M_{n+1}$. Thus, under mild assumptions, $f_T$ can also be taken as given. 

Assuming that the sample paths of $X$ are $\beta-$Hölder continuous, prediction intervals for the mean value of $\widehat Y_{n+1}$ given the $X_{n+1}(T_{n+1,m})$, $1\leq m \leq M_{n+1}$, 
can be directly built using Algorithm \ref{algo:pred-intervals} described in Section \ref{sec:noiseless-inf}, resulting in 
\begin{equation}\label{pred-int-pi}
\text{PI}_{1-\delta}=\left[\widehat Y_{n+1} + M_{n+1}^{-1/2 - \beta/d} q_{\delta/2}, \widehat Y_{n+1} + M_{n+1}^{-1/2 - \beta/d}q_{1 - \delta/2} \right].
\end{equation}
The simulation results in Section \ref{sec:numeric} show good coverage for this prediction interval, and illustrate that it is much shorter than the prediction interval based on the CLT and the Gaussian limit when the sample mean estimator is used instead.

\subsubsection{Case of noisy covariates}\label{sec:pred-noisy}
For simplicity, let $K=1$ in model \eqref{flm-model}. When error-in-variables are present, the discrete observations $X_i(T_{i,m})$ are given by
\begin{equation}\label{noisy_regZ}
	Z_{i,m} = X_i(T_{i,m}) + \sigma(T_{i,m})e_{i,m}, \qquad 1 \leq m \leq M_i, 1 \leq i \leq n,
\end{equation}
where the error terms $e_{i,m}$ are independent copies of a random variable $e$ with zero mean and unit variance. We assume that $X$, $T$ and the $M_i$'s are mutually independent. A feasible version of \eqref{pred-noiseless} is then
\begin{equation}
\widehat Y_{n+1} = \alpha_0 + \sum_{m=1}^{M_{n+1}} w_{M_{n+1},m} \phi(T_{n+1, m}), \quad \text{ with } \; \phi(T_{n+1, m}) = \frac{\alpha(T_{n+1, m}) Z_{n+1,m} }{f_T(T_{n+1, m})}.
\end{equation}
Let 
$$
\widehat \Sigma_{n+1} = \sum_{m=1}^{M_{n+1}} w_{M_{n+1},m} \frac{\alpha(T_{n+1, m})\sigma(T_{n+1, m})}{f_T(T_{n+1, m})}e_{n+1,m}.
$$ 
The prediction can then be decomposed as
\begin{equation}
\widehat Y_{n+1} = \widehat I(\varphi) + \widehat \Sigma_{n+1} + R_{n+1}, 
\end{equation}
where $R_{n+1} = \int_{\mathcal{T}} \alpha(t)X_{n+1}(t) dt - \widehat I(\varphi)$ is the remainder term resulting from the ideal, infeasible integral approximation $\widehat I(\varphi)$  constructed with $\varphi(T_{n+1, m}) =  \alpha(T_{n+1, m}) X_{n+1}(T_{n+1, m}) /f_T(T_{n+1, m})$. Following Section \ref{sec:inf-noisy-int}, a $(1 - \delta)-$level prediction interval for the mean value of $Y_{n+1}$ given the functional covariate  is  
\begin{equation}
\text{CI}_{1 - \delta} = \left[\widehat Y_{n+1} - z_{1 - \delta/2}  s_{M_{n+1}} , \widehat Y_{n+1} + z_{1 - \delta/2}  s_{M_{n+1}} \right],
\end{equation}
where $z_{\delta}$ is the $\delta$-quantile of the $\mathcal N(0,1)$  distribution and, following Proposition \ref{CLT-prop}, 
$$
s^2_{M_{n+1}} =  \sum_{m=1}^{M_{n+1}} w^2_{M_{n+1},m}\frac{\alpha^2 (T_{n+1, m} ) \sigma^2(T_{n+1, m})}{f^2_T(T_{n+1, m})},
$$
is the conditional variance of $\widehat Y_{n+1}$ given the functional covariate observations. 
Like for the density $f_T$, the conditional variance $\sigma^2(\cdot)$ of the measurement errors for the functional predictor, can be estimated nonparametrically using the learning set of functional data. See, e.g, \cite{WKP}. In the case $\mathcal T = [0,1]$, an alternative $\text{CI}_{1 - \delta}$ can be constructed using the expression of the limit of $s^2_{M_{n+1}}$ derived in Proposition \ref{CLT-prop}, that is 
\begin{equation}\label{CI_as}
s^2_{M_{n+1}} = (5/2)M^{-1} \mathbb{E}_M\left[\alpha^2(T)\sigma^2(T)/f^2_T(T) \right]\{1+o_{\mathbb P}(1)\}.
\end{equation}

\medskip

\begin{remark}
From the point of view of our approach, the conditional variance $\sigma^2(\cdot)$ can also depend on the sample path of $X$, see \eqref{ini_mod} and the fPCA example below. This would require a more refined  procedure for learning $\sigma^2(\cdot)$ from the learning set, or additional modeling assumptions about this conditional variance. The issue is not specific for our approach, and the problem of learning the conditional covariance is also expected to be encountered in the competing approaches to constructing prediction intervals. 
\end{remark}

\subsubsection{Extensions to other predictive models}
Although we focused our exposition on the functional linear model, the control neighbors approach similarly applies for more general functional regression models. A natural extension is the generalized functional linear model, of the form
\begin{equation}\label{glm_fr}
	Y = g\left(\alpha_0 + \langle \alpha, X \rangle  \right) + \epsilon,\quad \text{ with } \; \mathbb{E}(\epsilon\mid X) = 0  \; \text{ and } \; \mathbb{E}(\epsilon^2\mid X) = \sigma^2_{\epsilon}(X),
\end{equation}
where $g(\cdot)$ is a monotone,  invertible link function. For example, with a binary response $Y \in \{0, 1\}$, as is the case   in supervised classification, the link function can be the logit function $g(x) =1/(1 + \exp(-x))$. The   prediction intervals for the mean value of $Y_{n+1}$ in the regression model \eqref{glm_fr} when the values of $X$ are observed without noise, are simply obtained as the image through the monotone function $g(\cdot)$ of the prediction intervals in \eqref{pred-int-pi}.


\subsection{fPCA Scores}
Let $\mu(t) = \mathbb{E}[X(t)]\in \mathbb R, \forall t \in \mathcal{T}$, be the mean function. Functional principal component analysis (fPCA) involves estimating the eigen-elements $(\lambda_j, \psi_j)_{j \geq 1}$ that solves the integral equation 
\begin{equation}
	\int_{\mathcal{T}} \Gamma(s,t) \psi_j(t) dt = \lambda_j \psi_j(s), 
\end{equation}
where $\Gamma(s,t) = \mathbb{E}\left[\left\{X(s) - \mu(s) \right\}\left\{X(t) - \mu(t) \right\}\right]$ is the covariance function.  The observations come in the form of pairs $(Z_{i,m}, T_{i,m}) \in \mathbb{R} \times \mathcal{T}, 1 \leq i \leq n, 1 \leq m \leq M_i$,  generated according to \eqref{ini_mod}. The mean, the eigen-functions and the density $f_T$ are considered given. 

By definition, the fPCA scores for the $i$-th curve $X_i$ are given by 
\begin{equation}\label{scores-eq}
	\xi_{i,j} =  \langle X_i - \mu,\psi_j \rangle = \mathbb{E}\left[\frac{\left\{X_i(T) - \mu(T)\right\}\psi_j(T)}{f_T(T)} \mid X_i \right].
\end{equation}
Once more, a distinction is drawn between the noiseless case and the noisy case, which correspond to a null and a positive conditional variance function for the errors, respectively.

\subsubsection{Estimation and inference with noiseless functional data}

When $Z_{i,m}= X_i(T_{i,m})\in\mathbb R$, given $\mu$, $\{\psi_j\}_{j \geq 1}$ and $f_T$, the scores can be estimated by 
\begin{equation}
\widehat \xi_{i,j} = \sum_{m=1}^{M_i} w_{M_i, m} \varphi_j(T_{i,m}), \quad \text{ with } \; \varphi_j(T_{i,m}) = \frac{\left\{X_i(T_{i,m}) - \mu(T_{i,m}) \right\}\psi_j(T_{i,m})}{f_T(T_{i,m})}.
\end{equation}
Prediction intervals can similarly be built using Algorithm \ref{algo:pred-intervals}, leading  to the approximate $(1 - \delta)-$level interval 
\begin{equation}
\left[\widehat \xi_{i,j} + M_i^{-1/2-\beta/d}q_{\delta/2}, \;\; \widehat \xi_{i,j} + M_i^{-1/2-\beta/d}q_{1 - \delta/2} \right].
\end{equation}

\subsubsection{Case of noisy observations}
When the discrete observations $X_i(T_{i,m})\in\mathbb R$ are contaminated with noise, \emph{i.e.,} $\sigma^2(\cdot;\cdot)>0$ in \eqref{ini_mod}, 
a feasible estimate of the scores are  given by 
\begin{equation}
\widehat \xi_{i,j} = \sum_{m=1}^{M_i} w_{M_i, m} \phi_j(T_{i,m}), \quad \text{ with } \; \phi_j(T_{i,m}) = \frac{\left\{Z_{i,m} - \mu(T_{i,m}) \right\}\psi_j(T_{i,m} )}{f_T(T_{i,m})}.
\end{equation}
By Proposition \ref{CLT-prop}, the corresponding confidence intervals for $ \xi_{i,j} =  \langle X_i - \mu,\psi_j \rangle$ are then given by
\begin{equation}
\left[\widehat \xi_{i,j} - z_{1 - \delta/2} s_{M_i}, \;\; \widehat \xi_{i,j} + z_{1 - \delta/2}  s_{M_i} \right],
\end{equation}
where  
$$
s^2_{M_{i}} =  \sum_{m=1}^{M_{i}} w^2_{M_{i},m}\frac{ \sigma^2(T_{i, m},X_i(T_{i,m} ))\psi_j^2 (T_{i, m} )}{f^2_T(T_{i, m})}.
$$

\begin{remark}
We here considered the most popular basis in fPCA, that given by the eigen-functions of the covariance operator. Such a data-driven basis requires to be estimated. Alternative, the score calculation we propose can be considered with a fixed basis (Fourier, B-splines, \emph{etc}). 
\end{remark}


\subsection{Outlier detection by data depths} 

Data depth is an extension of the sample median to more general sample spaces than the real line. Let $P$ be the probability distribution of the vector-valued random function $X$, with $P(t)$ denoting the marginal probability of $X(t)\in\mathbb R^K$; $K\geq 1$. To a given sample path of $X$, a data depth assigns a non-negative number, interpreted as a measure of centrality of this sample path with respect to the probability distribution $P$. The existing approaches towards the assignment of a depth value to a random function can be categorized into two distinct families: integrated depths and non-integrated, or geometric depth. See \cite{Gerda2014}, \cite{NagyGijbels2016}, \cite{battey},  \cite{gij2017}. 

The integrated depths for vector-valued, multivariate (domain) functional data have a form of an integral 
\begin{equation}\label{eq:multivariate-fd}
MFD(x;P,D) = \int_{\mathcal{T}} D(x(t); P(t)) \Omega(t) dt, 
\end{equation}
where $\Omega(t)$ is an arbitrary, non-negative weight function integrating to 1. Different choices for the depth function $D$ in \eqref{eq:multivariate-fd} are available. See the reviews \cite{battey},  \cite{gij2017}.  Data depths serve as a useful tool in outlier detection, see \cite{Febreto2008}. 


Although most work on data depths in the FDA setting have focused on the common design framework, extensions to the random design case have been recently explored; see \cite{NagyGijbels2016} and \cite{Nagy2019}. Under the random design framework, the  functional depth of a sample path $X_i$ can be written as 
\begin{equation}
MFD(X_i; P, D) = \mathbb{E}\left[\frac{D(X_i(T); P(T)) \Omega(T)}{f_T(T)}  \mid X_i \right].
\end{equation}
In the noiseless case, \emph{i.e.,} the $ X_i(T_{i,m})\in\mathbb R$, $1\leq m\leq M_i$ are observed,  given $\Omega, D, P$ and $f_T$, the depths can be estimated by 
\begin{equation}
\widehat{MFD}_i = \widehat{MFD}(X_i; P, D) = \sum_{m=1}^{M_i} w_{M_i, m} \varphi(T_{i,m}), \qquad \text{with } \; \varphi(T_{i,m}) = \frac{D(X_i(T_{i,m}); P(T_{i,m})) \Omega(T_{i,m})}{f_T(T_{i,m})},
\end{equation}
and prediction intervals can be similarly built using Algorithm \ref{algo:pred-intervals}, with an approximate $(1-\delta)-$level interval given by
\begin{equation}\label{depthPI}
\left[\widehat{MFD}_i + M_i^{-1/2-\beta/d} q_{\delta/2},\;\; \widehat{MFD}_i + M_i^{-1/2 - \beta/d} q_{1 - \delta/2} \right].
\end{equation}

\medskip

\begin{remark}
The common functions $D$ in \eqref{eq:multivariate-fd} is Lipschitz continuous in the first arguments. Then, in the case of differentiable sample paths $X_i$, the value of $\beta$ for the prediction interval \eqref{depthPI} is equal to 1. On contrary, with non-differentiable $X_i$, the value of $\beta$ is given by the regularity of the sample path. See also Section \ref{disc_beta} for a discussion. 
	
\end{remark}


\section{Random integrands}\label{disc_beta}
Since in the FDA context the integrand $\varphi(t,X(t))$ depends on the sample path of $X$, its regularity is a subtle issue. 
In all the application examples considered above, the map $(t,x)\mapsto \varphi(t,x)$ is smooth, that is it admits at least continuous first-order partial derivatives on $\mathcal T \times \mathbb R$. Then, the regularity parameter $\beta$ is determined by the regularity of the sample paths of the process $X$. In the FDA literature it is  quite often supposed  that the sample paths of $X$ are continuously differentiable. In this case we have $\beta=1$ in our approach and there is no issue for the practitioner as to how to set the value of $\beta$.

Recently, the case where the sample paths are non-differentiable has received much attention. There is now extensive evidence that in some applications, such as energy and climate, chemistry and physics, sports science and medical applications, many functional data sets can reasonably be assumed to be generated by continuous but irregular sample paths of $X$. 
See, for example,   \cite{impact20}, \cite{Petrovich_irreg}, \cite{Mohammadi2021}, \cite{Mohammadi2022}, \cite{WKP}. Typically, the paths can reasonably be assumed to be Hölder continuous, but the Hölder exponent is generally unknown.  In the case of non-differentiable sample paths observed without error at random design points, the choice of the  Hölder exponent $\beta$ is a subtle issue that affects the convergence rate of the control neighbor estimate and the scaling factor in the subsampling procedure in Algorithm \ref{algo:pred-intervals}. Fortunately, probability theory and recent contributions to the FDA literature provide some guidance.

Consider the class of zero-mean processes $X$ for which positive constants $\zeta$, $\kappa$, $C_X$ exists 
such that
\begin{equation}\label{sub-gauss-incre}
\mathbb{E}\left( |X(t) - X(s)|^\zeta \right) \leq C_X d(t,s)^{d + \kappa}, \qquad \forall t, s \in \mathcal{T}\subset \mathbb R^d.
\end{equation}
If the process $X$ satisfies \eqref{sub-gauss-incre}, then the Kolmogorov-Chentsov continuity theorem states there exists a Hölder continuous modification such that $X$ is $\gamma$-Hölder continuous for all $ 0 < \gamma < \kappa/\zeta$. See, e.g., \citet[Chapter I, Theorem 2.1]{Yor1999}, \cite{urusov23}. Then, a question to be answered is what is the value of $\beta$ for processes satisfying \eqref{sub-gauss-incre}. An answer is given by the recent contributions \cite{golo1} and \cite{WKP} in the case $\mathcal T = [0,1]$, and \cite{kassi2023} when $\mathcal T = [0,1]^2 $. See also \cite{hsing2016}, \cite{shentsing2020} for related problems. In the case of design point in the unit interval on the real line, the idea is based on the remark that for many zero-mean processes $X$ with non-differentiable sample paths, it holds that, for sufficiently small $\delta >0$, 
\begin{equation}\label{local_reg}
\mathbb{E}\left( |X(t+\delta/2) - X(t-\delta/2)|^2 \right) \approx L_t^2 \delta^{2H_t},
\end{equation}
where $t\mapsto H_t \in(0,1)$ and $t\mapsto L_t >0$ are continuous functions. The functions $H$ and $L$ characterize the local regularity of the sample paths of $X$. The smaller the $H_t$, the more irreggular the paths are. \cite{WKP} provide examples of a large class of Gaussian processes, including the fractional Brownian motion. The class can be extended by several types of transformations. By suitable moment conditions for 
$|X(t+\delta/2) - X(t-\delta/2)|\delta^{-\underline H}$, with $\underline H = \min_{t\in\mathcal T} H_t$, it is the possible to check \eqref{sub-gauss-incre} with $d+\kappa = \zeta \underline H$, for any $\zeta \geq 2$. By the Kolmogorov-Chentsov continuity theorem,  it then follows that there exists a Hölder continuous modification such that $X$ is $\beta$-Hölder continuous for all $ 0 < \beta < \underline H$. As example with $d=1$, in the Brownian motion case,  $H_t$ is constant equal to 1/2, and the sample path are $\beta-$Hölder continuous for any $\beta<1/2$. 

The function $H_t$ in \eqref{sub-gauss-incre} can be learned from the learning data set. In the case $\mathcal T = [0,1]$, \cite{WKP} derived exponential bounds for the uniform concentration of the estimator  $\widehat H_t$ of $H_t$, from which the bounds for the concentration of $\widehat {\underline H}=\min_t \widehat H_t$, the estimator of $\underline H$, can be derived. In particular, it was shown that  the concentration rate of $\widehat {\underline H}$ is faster than any negative power of $\log(M)$. On the basis of the concentration results for $\widehat {\underline H}$, and the fact that $M^{1/\log^{a}(M)} \rightarrow 1$ provided that $a>1$, a sensible choice will then be to define the estimate of  $\beta$ as
\begin{equation}\label{beta-choice}
	\widehat \beta = \widehat {\underline H} - \log^{-2}(M).
\end{equation}
where the extra log term corresponds to the rate of covergence of $\widehat \beta$. A detailed theoretical analysis of the properties of the choice \eqref{beta-choice} is beyond  the scope of this paper.



\section{Numerical Results}\label{sec:numeric}
In this section, we explore the finite sample properties of the proposed estimates and inference procedures. We will focus on univariate functional data in the functional linear regression case, before moving on to multivariate functional data for fPCA scores.

\subsection{Linear functional regression}\label{sec:regression-simu}
In order to isolate the error stemming from integral estimation, we treat the intercept $\alpha_0$, slope function $\alpha$ and density $f_T$ as given quantities. Let $e_k$ be the eigenfunctions of the standard Brownian motion (Bm), given by
\begin{equation}
e_k(t) = \sqrt{2}\sin\left(k - 1/2\right)\pi t, \qquad \forall t \in \mathcal{T}=[0,1],\; k\geq 1.
\end{equation}
The online sample path $X_{n+1}$ is simulated using the truncated Kosambi-Karhuen-Loève decomposition
\begin{equation}\label{kl-true}
X_{n+1}(t) =\sum_{k=1}^K \xi_{n+1,k} e_k(t), \qquad \forall t \in \mathcal{T},
\end{equation}
where $\xi_{n+1,k}$ are the scores, given by
\begin{equation}
\xi_{n+1,k} = Z_{n+1,k} \sqrt{\lambda_k(\nu)}, \quad \text{ with } \quad \lambda_k(\nu) = (k - 1/2)^{-\nu}\pi^{-\nu}, \quad Z_{n+1,k} \sim \mathcal{N}(0, 1).
\end{equation}
Here, $\lambda_k(\nu)$ are the eigenvalues whose rate of decay can be adjusted by the parameter $\nu$. The eigenvalues of the standard Bm correspond to $\nu=2$. A faster rate of decay corresponds to a larger regularity $H$, which in this context is constant. If $X$ is represented as in \eqref{kl-true} with $K=\infty$, it can be shown the 
$\nu = 1 + 2H$ if $0<\nu<3$ and $H\in (0,1)$ in \eqref{local_reg}.
The intercept and slope function was taken to be 
\begin{equation}\label{slope-true}
\alpha_0 = 0, \quad \text{ and } \quad \alpha(t) = \sum_{k=1}^K 4(-1)^{k+1} k^{-p} e_k(t),
\end{equation}
a similar setup to \cite{caihall2006}.  A plot of the true slope  $\alpha$ is given in the Supplement, and we can see that it is almost linear. Using the orthonomal eigen-functions   in \eqref{slope-true} allows us to obtain an exact expression for the best linear prediction for the mean value of $Y_{n+1}$, see \eqref{infeasible-reg-pred}, of the form
\begin{equation}
\widetilde Y_{n+1} = \alpha_0 + \langle X_{n+1} ,\alpha\rangle = \sum_{k=1}^K 4(-1)^{k+1} k^{-p} \xi_{n+1,k} . 
\end{equation}


The sample paths $X_{n+1}$ were built with $K = 50$ basis functions. A range of rates $\nu \in \{2, 3, 4\}$, with $\nu = 2$ and $\nu=3$ is considered, corresponding to the Brownian motion and Lipschitz continuous sample paths, respectively. Although no further gains in convergence rate is made in theory with $\nu > 3$, we decided to include a higher order smoothness to explore the finite sample properties. The observed design points $\mathbf{T} = (T_1, \dots, T_M)$ were generated with the density $f_T(t) = 1 - b/2 + bt$ using inverse transform sampling on $\mathcal{T} = [0, 1]$, with $b \in \{0, 0.5\}$, including thus the uniform design case. The sample sizes were set to $M_i = M \in \{50, 100, 200\}, 1 \leq i \leq n$. A number of 2000 replications were performed on all combinations of $M, b$ and $\nu$, resulting in 18 different configurations. 

The relative estimation error is reported on the log scale, with zero indicating equal performance. They are defined as 
\begin{equation}\label{risk-int-ratio}
\mathcal{R}\left(\widehat I^c(\varphi), \widehat I(\varphi)\right) = \log\left( \left|\widehat I(\varphi) - I(\varphi) \right| \right)  - \log\left( \left|\widehat I^c(\varphi) -  I(\varphi) \right|\right),
\end{equation}
where $\widehat I^c(\varphi)$ is the integral estimator of a competing method. Comparisons are made to the trapezoidal rule, denoted $\widehat I^{\rm trapez}(\varphi)$, and the sample mean, denoted $\widehat I^{\rm mean}(\varphi)$. The latter is a frequently used estimator in the context of regression, see for example \cite{crambespline}.

\begin{figure}[!ht]
\centering
\includegraphics[height = 0.2\textheight,width=.18\textwidth]{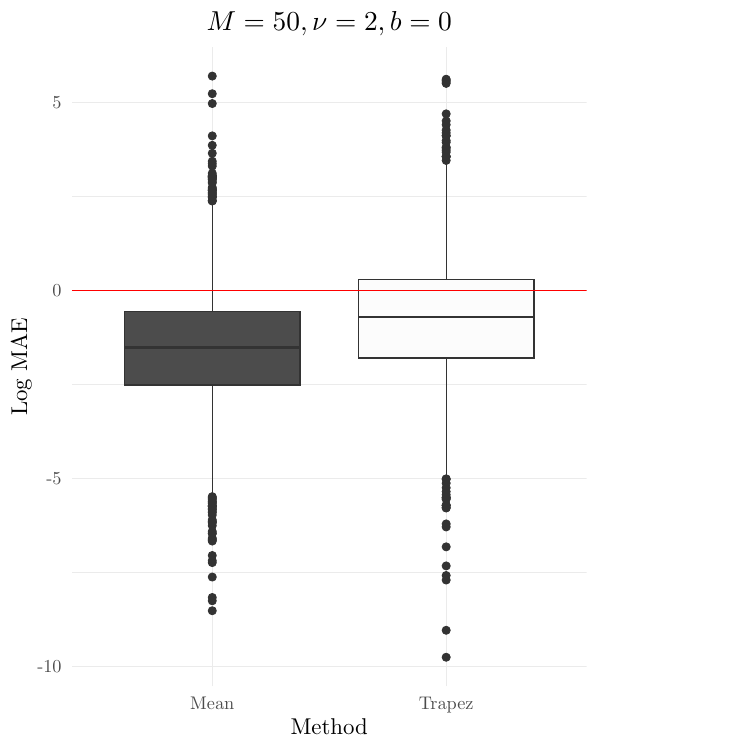}
\!\!\!\!\!\!\!\!\includegraphics[height = 0.2\textheight,width=.18\textwidth]{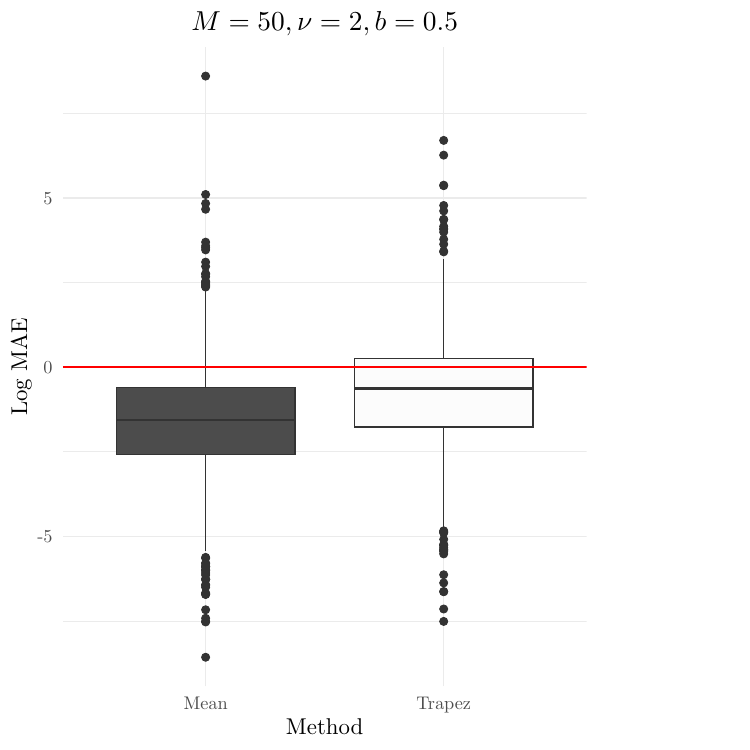}
\!\!\!\!\!\!\!\!\includegraphics[height = 0.2\textheight,width=.18\textwidth]{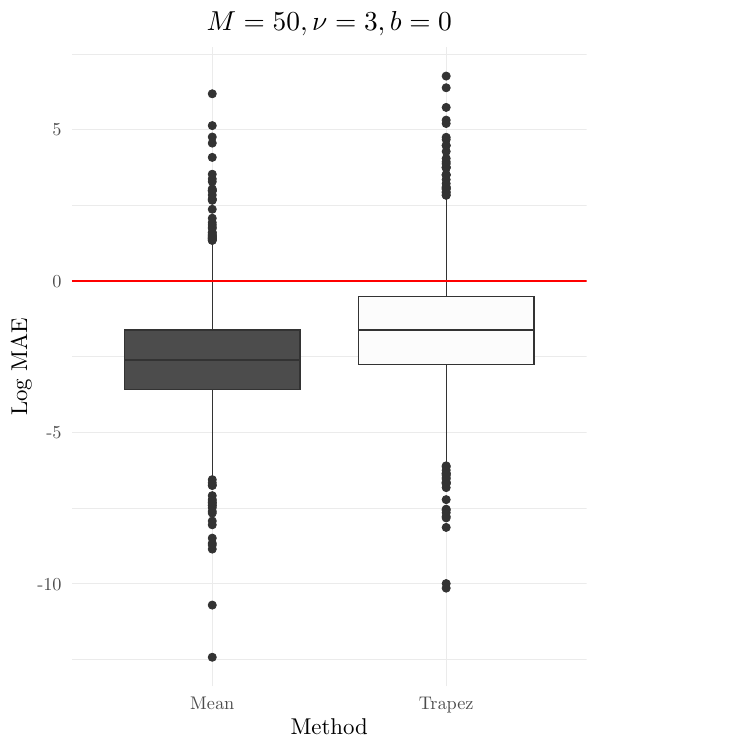}
\!\!\!\!\!\!\!\!\includegraphics[height = 0.2\textheight,width=.18\textwidth]{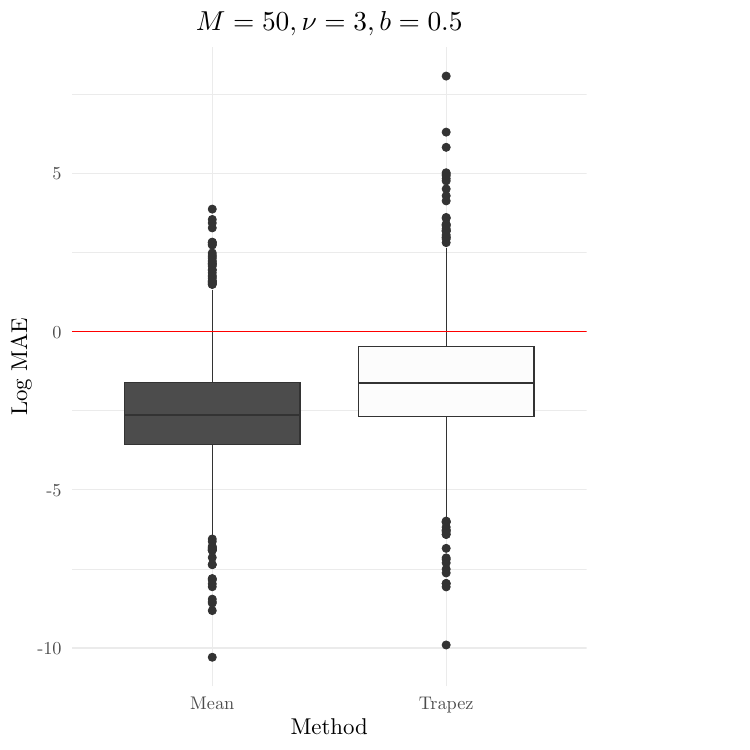}
\!\!\!\!\!\!\!\!\includegraphics[height = 0.2\textheight,width=.18\textwidth]{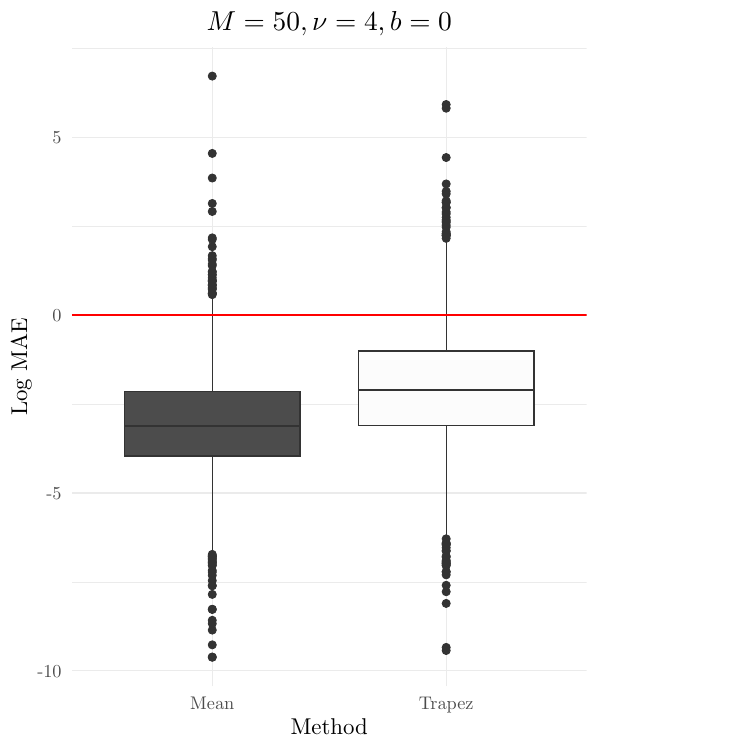}
\!\!\!\!\!\!\!\!\includegraphics[height = 0.2\textheight,width=.18\textwidth]{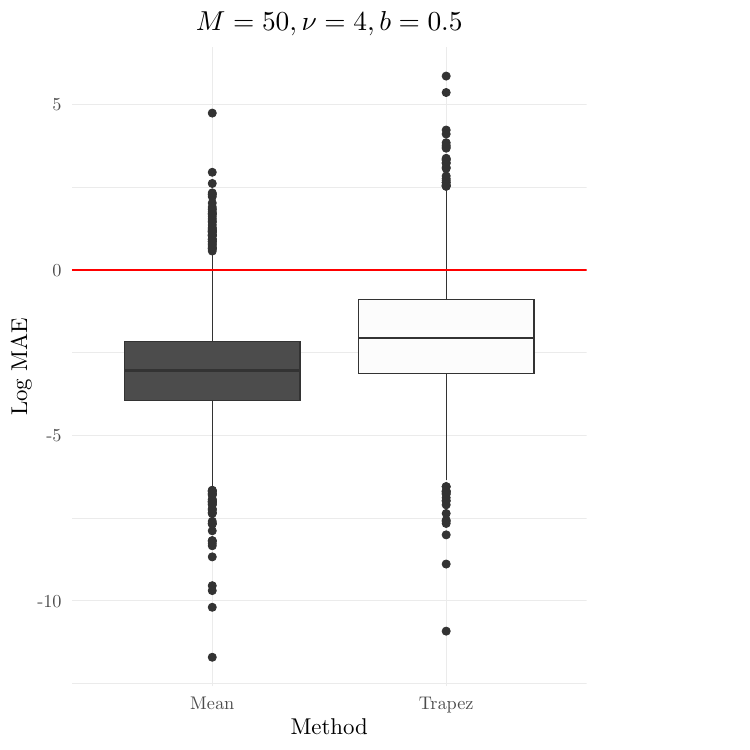}

\!\!\!\!\!\!\!\!\includegraphics[height = 0.2\textheight,width=.18\textwidth]{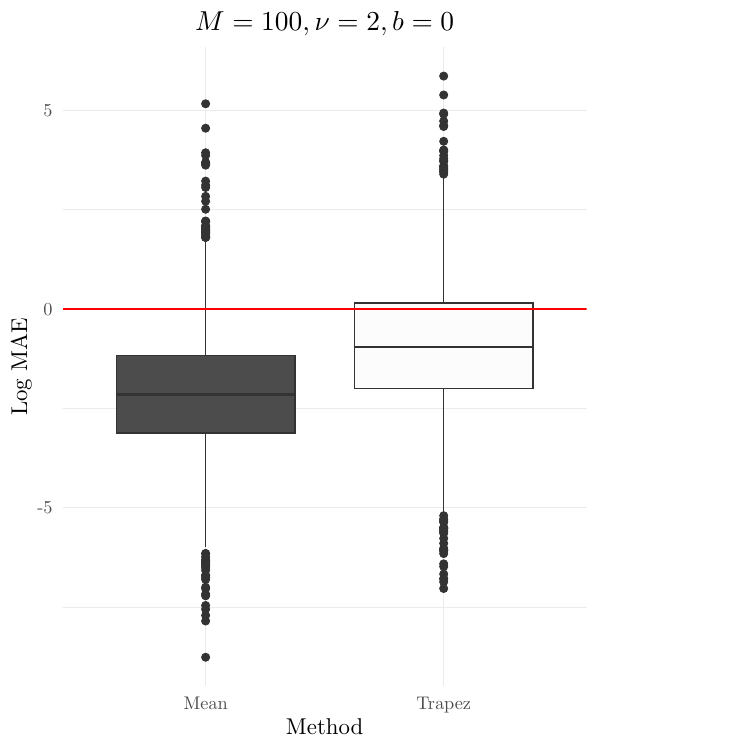}
\!\!\!\!\!\!\!\!\includegraphics[height = 0.2\textheight,width=.18\textwidth]{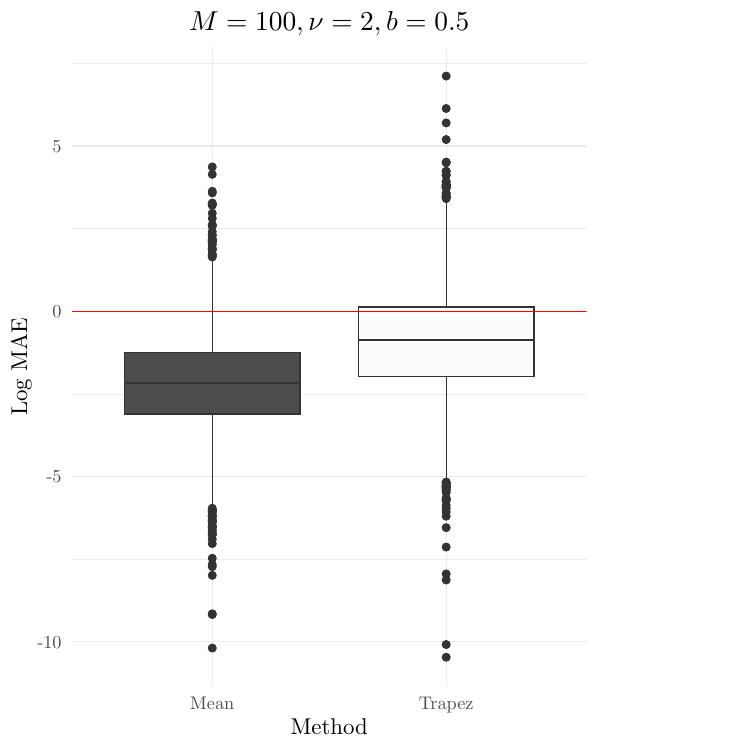}
\!\!\!\!\!\!\!\!\includegraphics[height = 0.2\textheight,width=.18\textwidth]{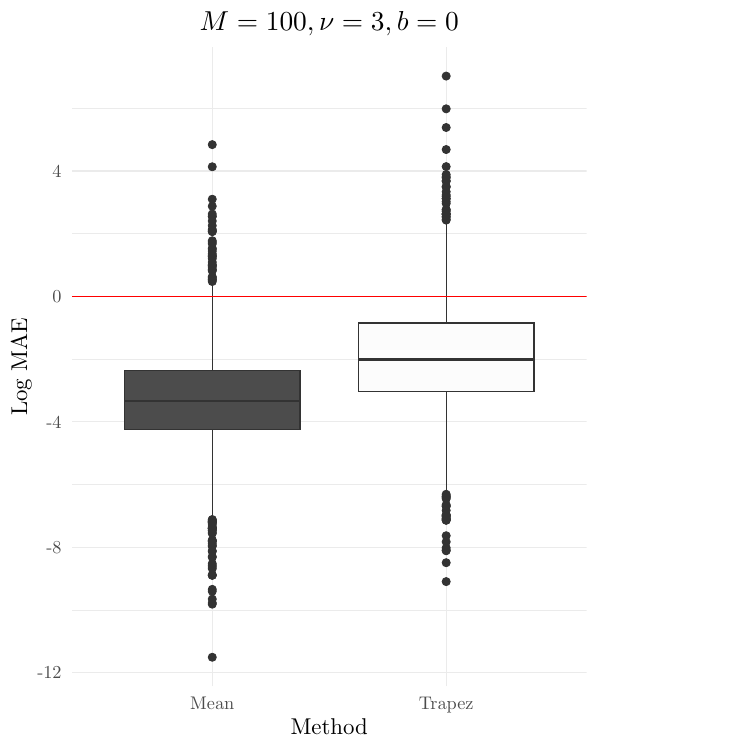}
\!\!\!\!\!\!\!\!\includegraphics[height = 0.2\textheight,width=.18\textwidth]{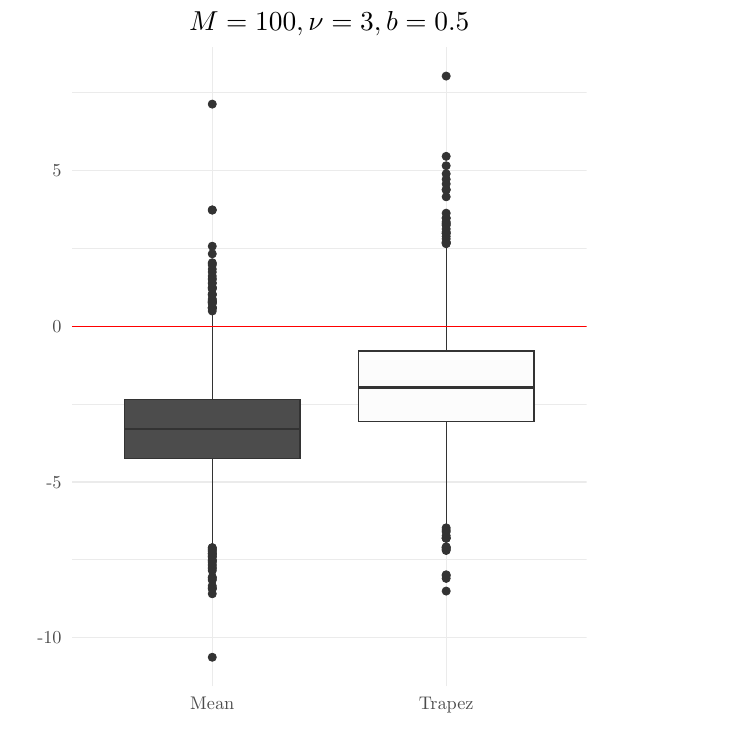}
\!\!\!\!\!\!\!\!\includegraphics[height = 0.2\textheight,width=.18\textwidth]{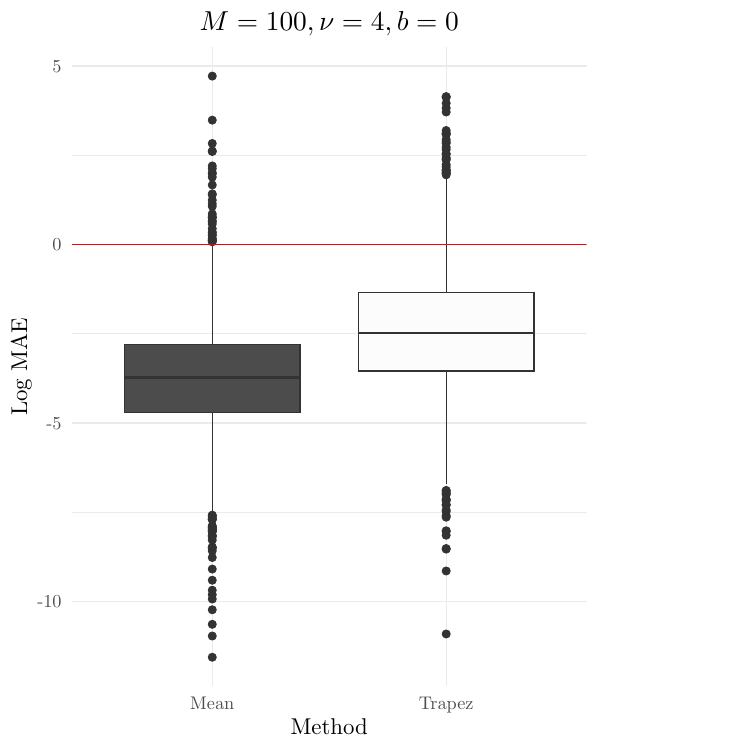}
\!\!\!\!\!\!\!\!\includegraphics[height = 0.2\textheight,width=.18\textwidth]{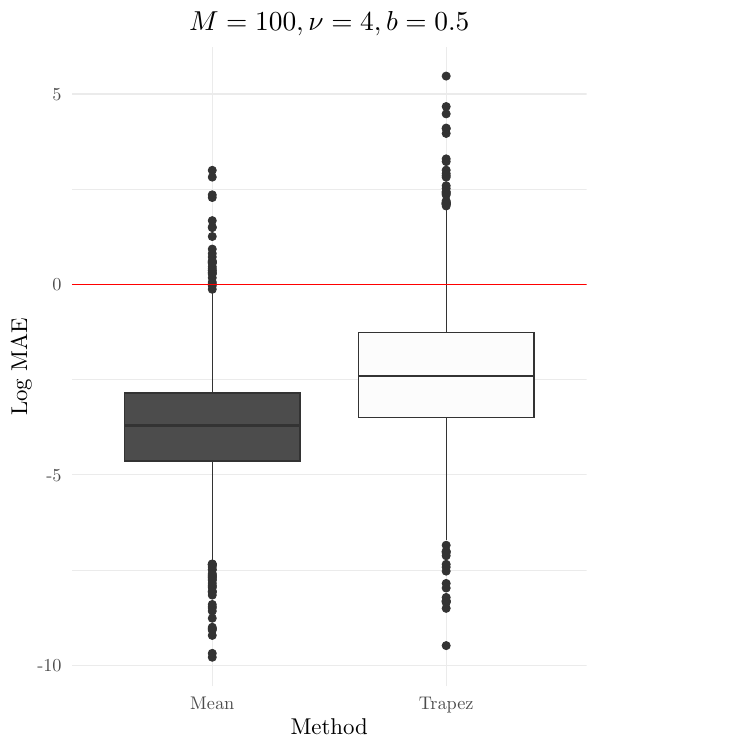}

\!\!\!\!\!\!\!\!\includegraphics[height = 0.2\textheight,width=.18\textwidth]{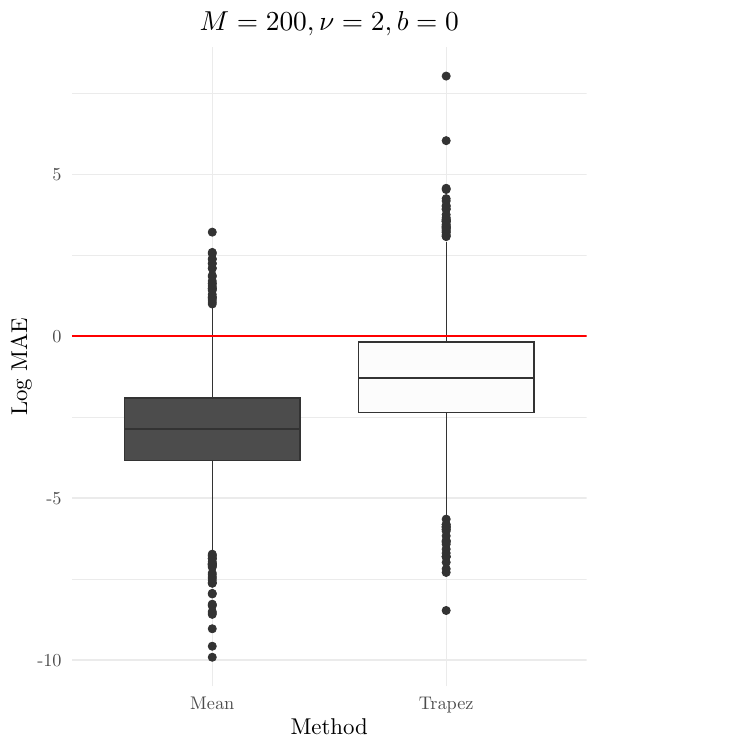}
\!\!\!\!\!\!\!\!\includegraphics[height = 0.2\textheight,width=.18\textwidth]{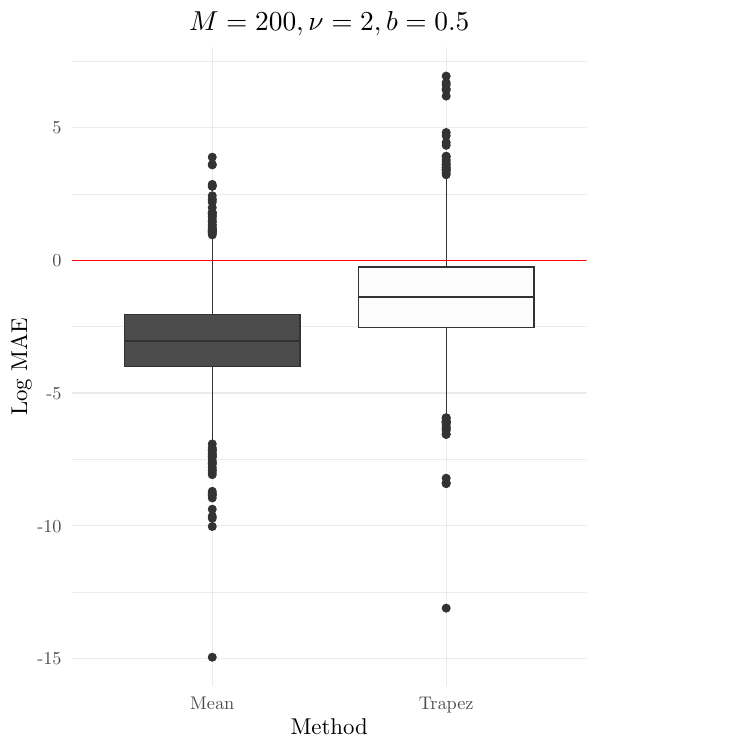}
\!\!\!\!\!\!\!\!\includegraphics[height = 0.2\textheight,width=.18\textwidth]{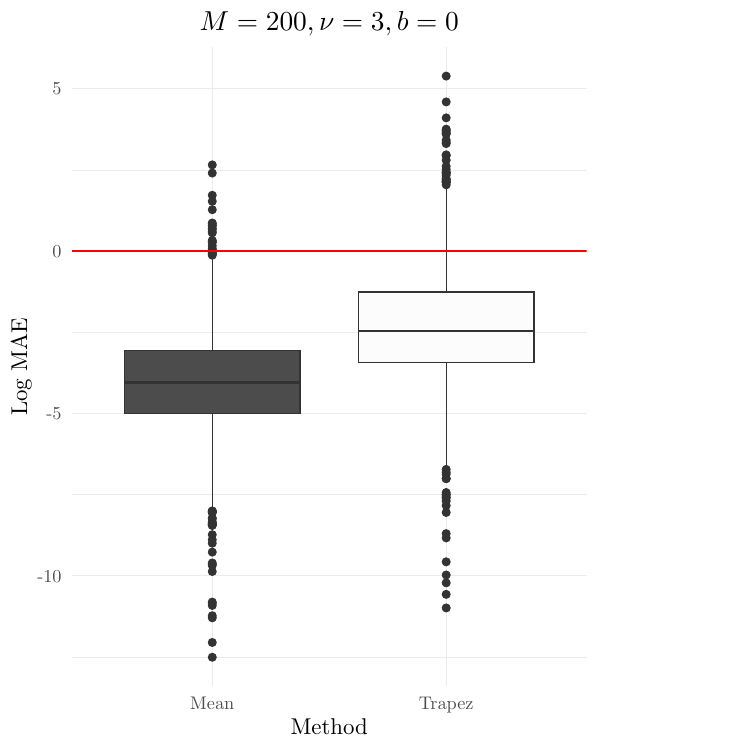}
\!\!\!\!\!\!\!\!\includegraphics[height = 0.2\textheight,width=.18\textwidth]{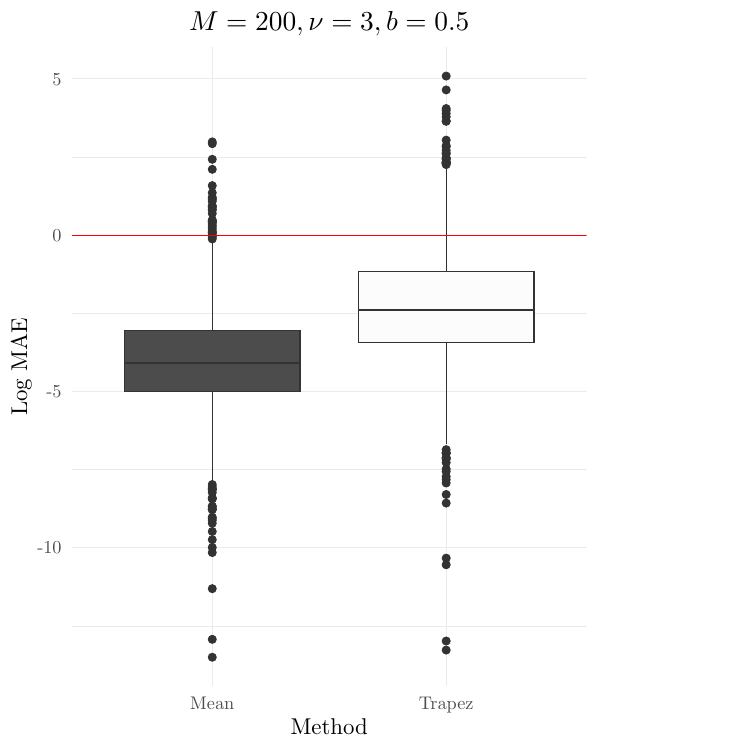}
\!\!\!\!\!\!\!\!\includegraphics[height = 0.2\textheight,width=.18\textwidth]{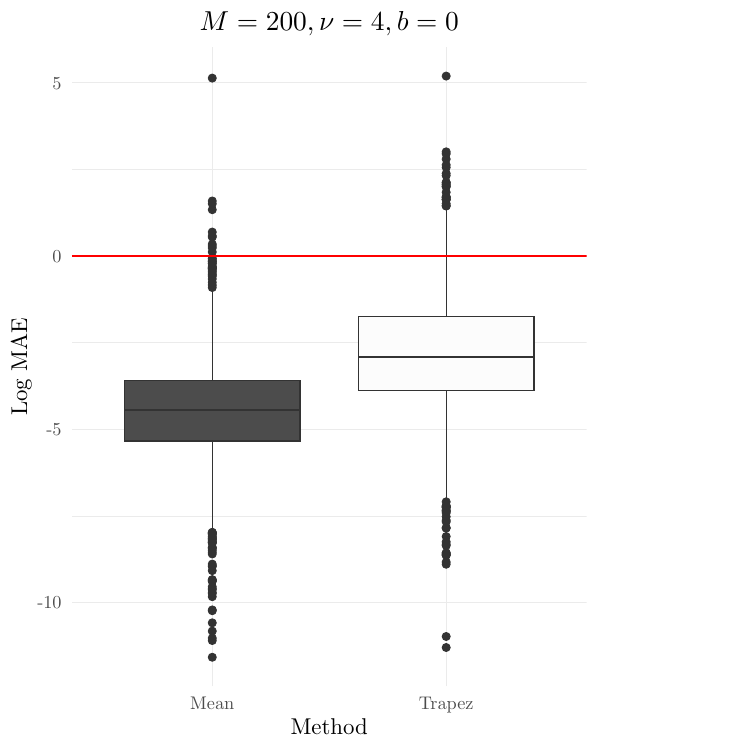}
\!\!\!\!\!\!\!\!\includegraphics[height = 0.2\textheight,width=.18\textwidth]{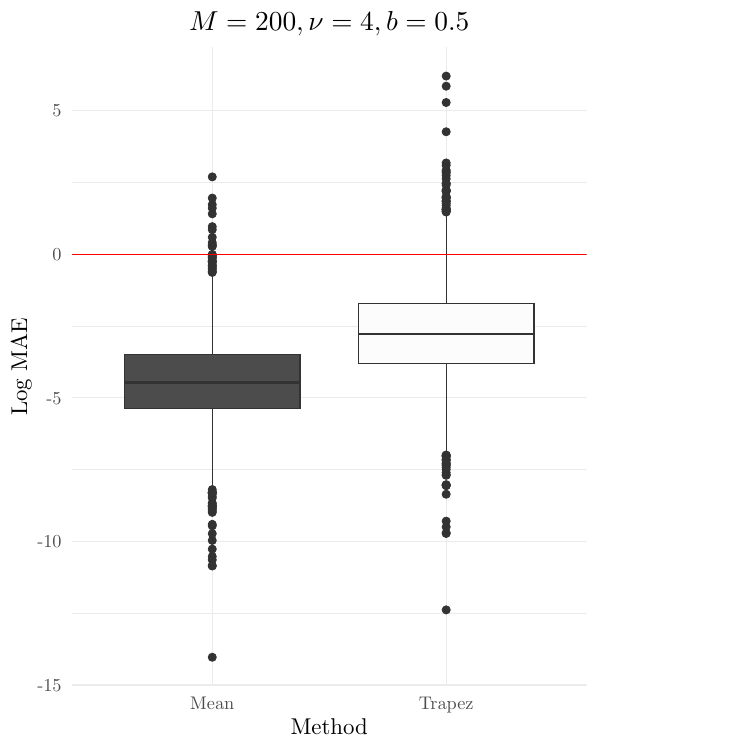}

\caption{\small Best linear prediction in  linear functional regression with noiseless covariate $X$: boxplots from 2000 replications of the log-ratios of absolute errors $\mathcal{R}(\widehat I^c(\varphi), \widehat I(\varphi))$  for $c\in\{{\rm mean, trapez}\}$ and different configurations: sample paths generated according to \eqref{kl-true} with $\nu\in\{2,3,4\}$, the design points density is $f_T(t) = 1 - b/2 + bt$ with $b\in\{0,0.5\}$, and $M\in\{50,100,200\}$. Comparison results below the zero level indicate a better performance for the control neighbors estimates.}
\label{fig:box-reg-noiseless}
\end{figure}

We first consider the noiseless covariate, that is the values $X_{n+1}(T_m)$ are observed. Boxplots displaying the logarithms of the prediction error ratios, as defined  in \eqref{risk-int-ratio}, are provided in Figure \ref{fig:box-reg-noiseless}. We see that the control neighbor methods performs significantly better than the competing ones.

Related to the prediction intervals with noiseless covariate, theoretical results on the convergence in distribution for the $\widehat I^{\rm trapez}(\varphi)$ are not easily available. For this reason, and for the sake of fair comparisons, the subsampling approach will also be used for the competitors {\rm `mean', `trapez'}. For the sample mean, comparisons can be made to both subsampling and the Gaussian limit given by the CLT. In the latter approach, the theoretical variance is replaced by its empirical counterpart. Denote the coverage levels of the competing prediction intervals with $p^{c}, c \in \{{\rm NN, trapez, m, ms}\}$, where `ms' refers to the sample mean prediction interval constructed with subsampling, and let $\ell^{c}$  be their lengths. We report $p^c$ and $\ell^c$ in Table \ref{tab:cov-width-reg-noiseless} for $1 - \delta = 0.95$. The lengths are averaged over the replications. We see that despite providing the best coverage, the control neighbor estimates have by far the shortest lengths. 

\begin{table}[ht]
\centering
\begin{tabular}{rrrrrrrrrrr}
  \hline
$M$ & $\nu$ & $b$ & $p^{\rm m}$ & $p^{\rm NN}$ & $p^{\rm trapez}$ & $p^{\rm ms}$ & $\ell^{\rm m}$ & $\ell^{\rm NN}$ & $\ell^{\rm trapez} $ & $\ell^{\rm ms}$ \\ 
  \hline
   50 &   2 & 0.0 & 0.94 & 0.94 & 0.53 & 0.81 & 1.31 & 0.27 & 0.48 & 0.91 \\ 
   50 &   2 & 0.5 & 0.94 & 0.95 & 0.58 & 0.80 & 1.13 & 0.22 & 0.42 & 0.79 \\ 
   50 &   3 & 0.0 & 0.93 & 0.94 & 0.24 & 0.81 & 0.78 & 0.05 & 0.18 & 0.55 \\ 
   50 &   3 & 0.5 & 0.93 & 0.95 & 0.27 & 0.80 & 0.66 & 0.04 & 0.15 & 0.46 \\ 
   50 &   4 & 0.0 & 0.94 & 0.97 & 0.12 & 0.82 & 0.56 & 0.02 & 0.12 & 0.39 \\ 
   50 &   4 & 0.5 & 0.93 & 0.97 & 0.15 & 0.81 & 0.46 & 0.02 & 0.10 & 0.32 \\ \hline
  100 &   2 & 0.0 & 0.95 & 0.97 & 0.54 & 0.82 & 0.91 & 0.12 & 0.27 & 0.64 \\ 
  100 &   2 & 0.5 & 0.95 & 0.97 & 0.57 & 0.83 & 0.79 & 0.10 & 0.23 & 0.55 \\ 
  100 &   3 & 0.0 & 0.94 & 0.96 & 0.20 & 0.82 & 0.55 & 0.02 & 0.10 & 0.38 \\ 
  100 &   3 & 0.5 & 0.94 & 0.96 & 0.22 & 0.82 & 0.46 & 0.01 & 0.08 & 0.32 \\ 
  100 &   4 & 0.0 & 0.94 & 0.98 & 0.07 & 0.82 & 0.39 & 0.01 & 0.07 & 0.27 \\ 
  100 &   4 & 0.5 & 0.94 & 0.98 & 0.10 & 0.82 & 0.32 & 0.01 & 0.05 & 0.23 \\ \hline
  200 &   2 & 0.0 & 0.95 & 0.99 & 0.50 & 0.83 & 0.64 & 0.05 & 0.14 & 0.45 \\ 
  200 &   2 & 0.5 & 0.95 & 0.99 & 0.51 & 0.83 & 0.56 & 0.04 & 0.11 & 0.39 \\ 
  200 &   3 & 0.0 & 0.95 & 0.98 & 0.14 & 0.83 & 0.39 & 0.01 & 0.05 & 0.27 \\ 
  200 &   3 & 0.5 & 0.95 & 0.98 & 0.15 & 0.82 & 0.32 & 0.005 & 0.04 & 0.23 \\ 
  200 &   4 & 0.0 & 0.95 & 0.99 & 0.03 & 0.83 & 0.28 & 0.003 & 0.03 & 0.19 \\ 
  200 &   4 & 0.5 & 0.95 & 0.98 & 0.04 & 0.83 & 0.23 & 0.0024 & 0.027 & 0.16 \\ 
   \hline
\end{tabular}
\caption{\small Coverage and average length of the prediction intervals
in linear functional regression with noiseless covariate $X$, with nominal coverage level $1 - \delta = 0.95$. 1000 subsamples were drawn in each of the 2000 replications. Comparisons made to $c \in \{{\rm trapez, m, ms}\}$, denoting to the trapezoidal rule, sample mean and sample mean with subsampling, respectively. The setups for generating  sample paths and design points are the same as for Fig.  \ref{fig:box-reg-noiseless}.} 
\label{tab:cov-width-reg-noiseless}
\end{table}

\begin{figure}[!ht]
\centering
\includegraphics[height = 0.2\textheight,width=.18\textwidth]{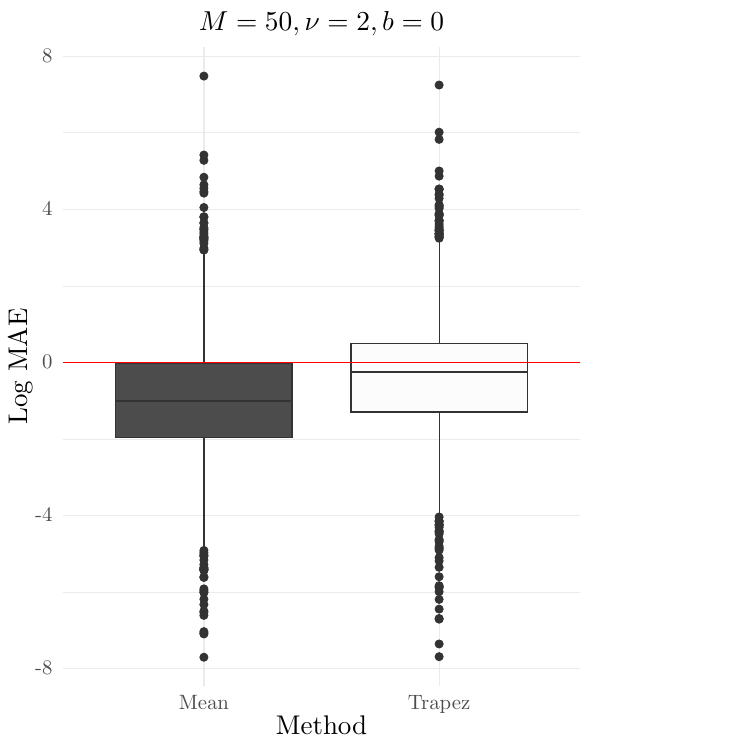}
\!\!\!\!\!\!\!\!\includegraphics[height = 0.2\textheight,width=.18\textwidth]{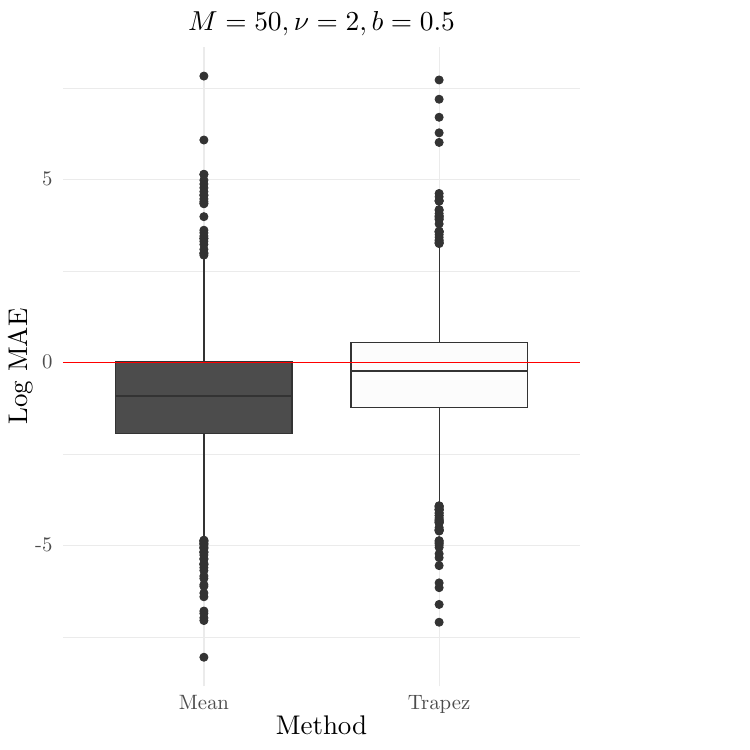}
\!\!\!\!\!\!\!\!\includegraphics[height = 0.2\textheight,width=.18\textwidth]{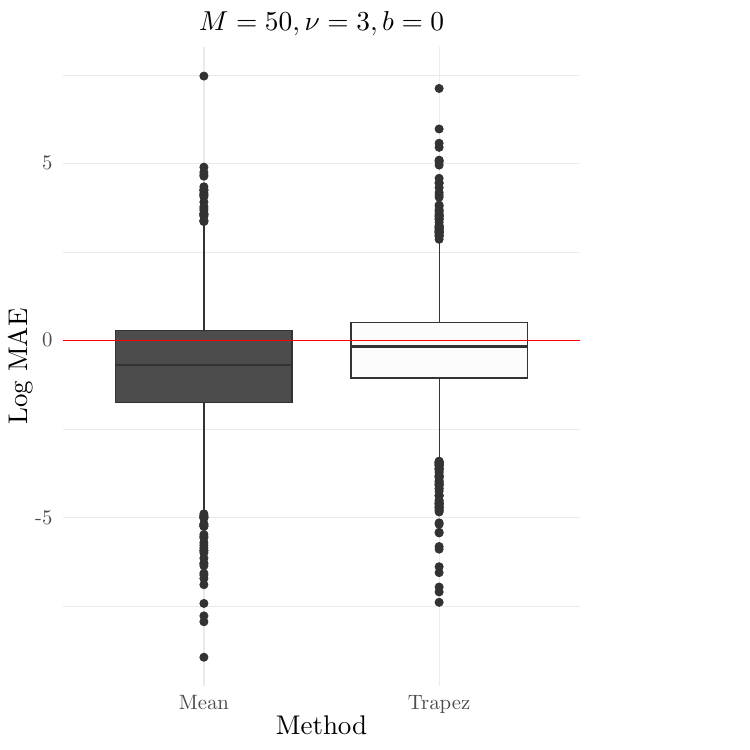}
\!\!\!\!\!\!\!\!\includegraphics[height = 0.2\textheight,width=.18\textwidth]{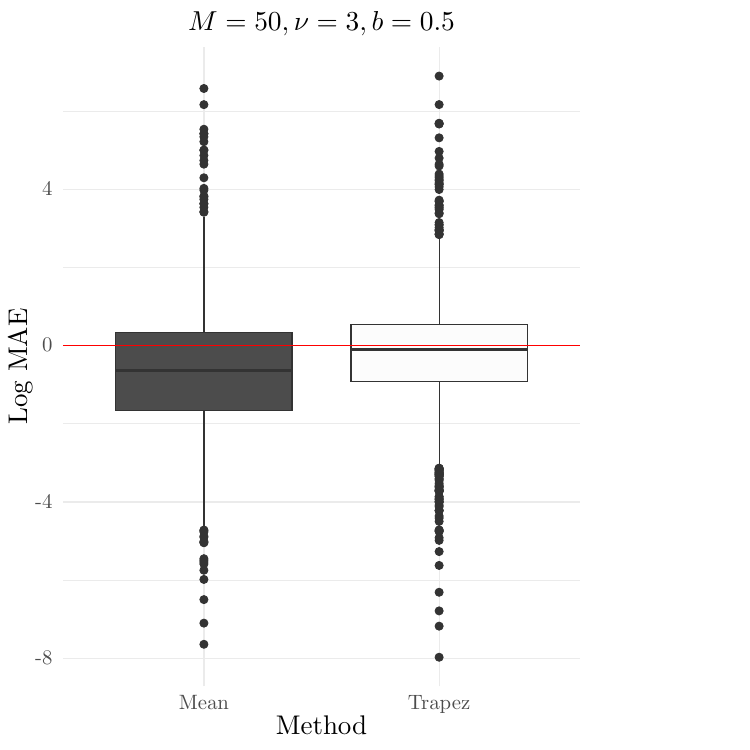}
\!\!\!\!\!\!\!\!\includegraphics[height = 0.2\textheight,width=.18\textwidth]{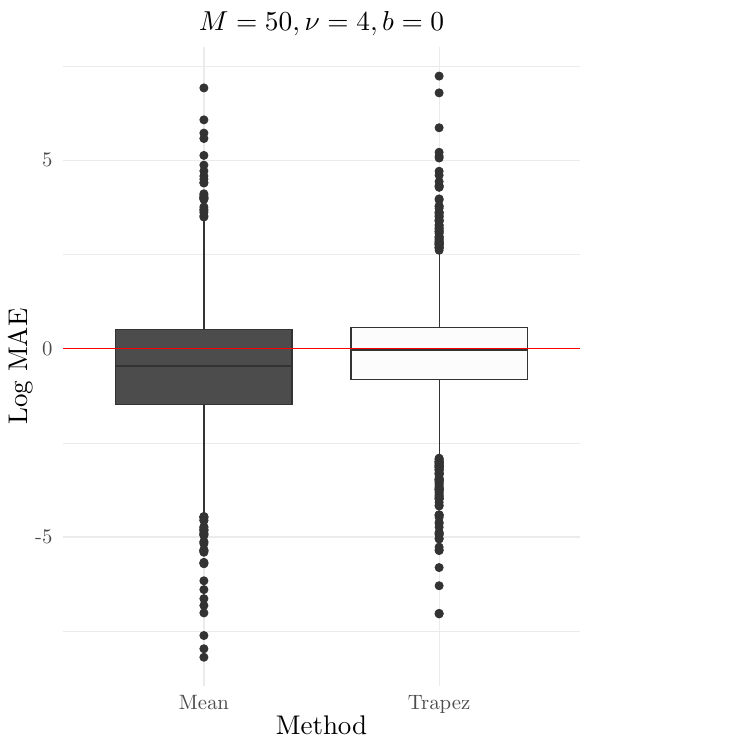}
\!\!\!\!\!\!\!\!\includegraphics[height = 0.2\textheight,width=.18\textwidth]{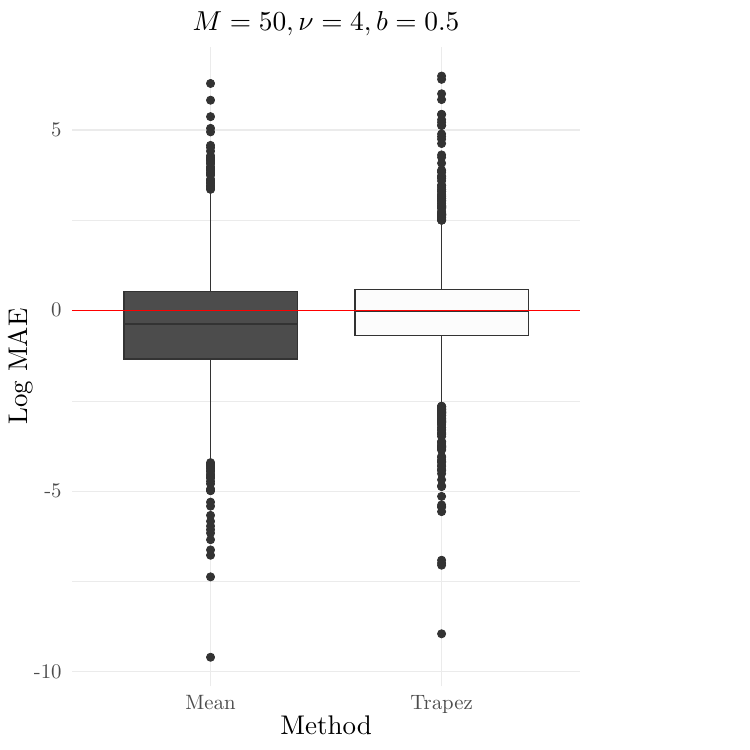}

\includegraphics[height = 0.2\textheight,width=.18\textwidth]{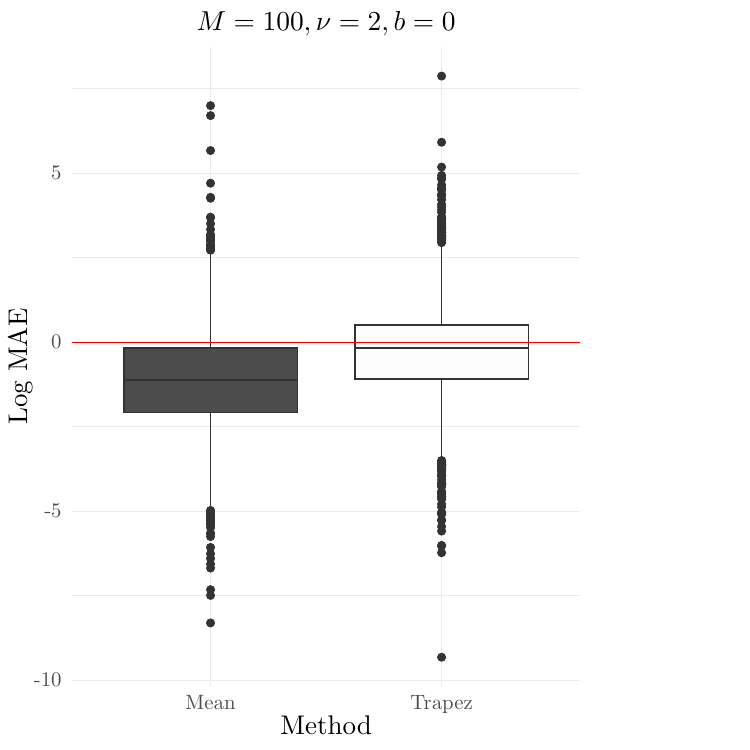}
\!\!\!\!\!\!\!\!\includegraphics[height = 0.2\textheight,width=.18\textwidth]{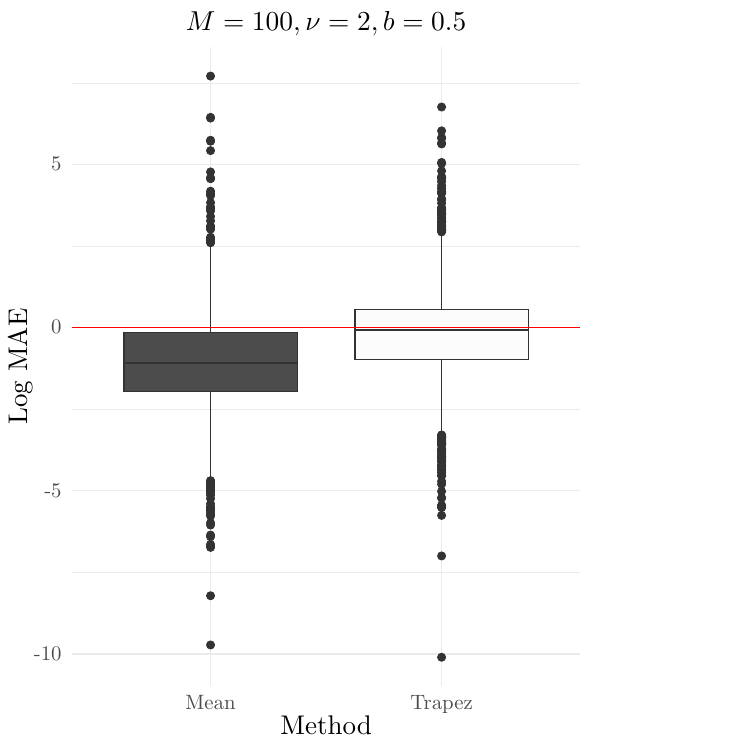}
\!\!\!\!\!\!\!\!\includegraphics[height = 0.2\textheight,width=.18\textwidth]{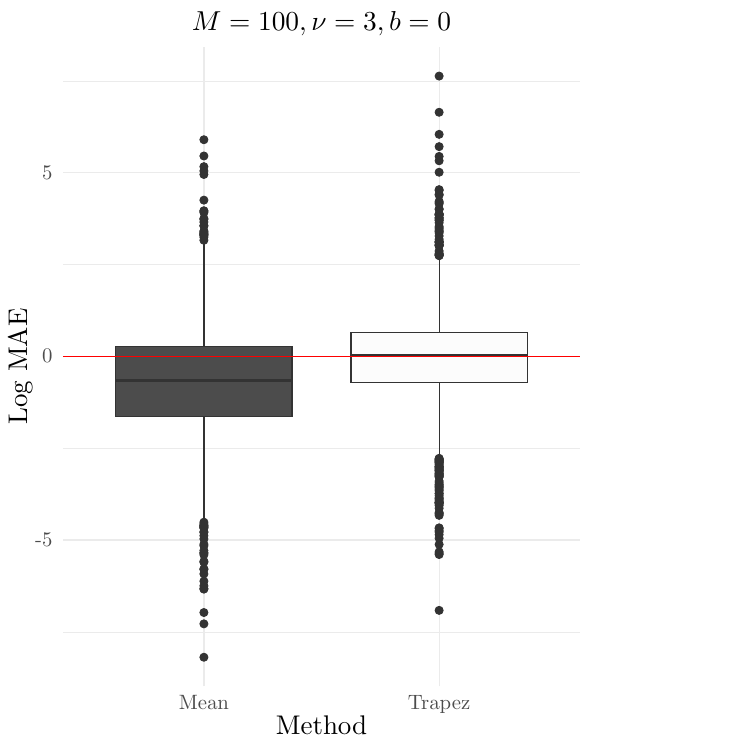}
\!\!\!\!\!\!\!\!\includegraphics[height = 0.2\textheight,width=.18\textwidth]{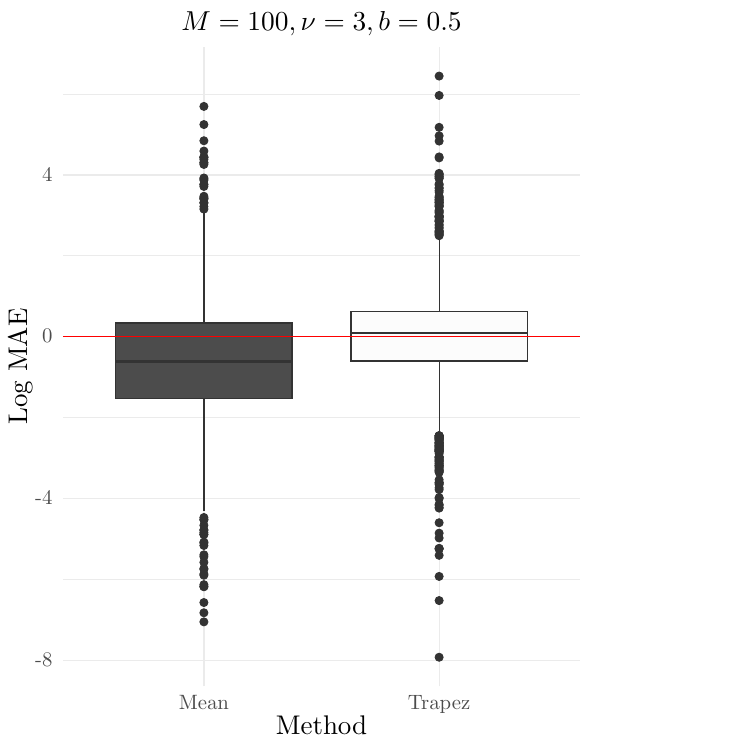}
\!\!\!\!\!\!\!\!\includegraphics[height = 0.2\textheight,width=.18\textwidth]{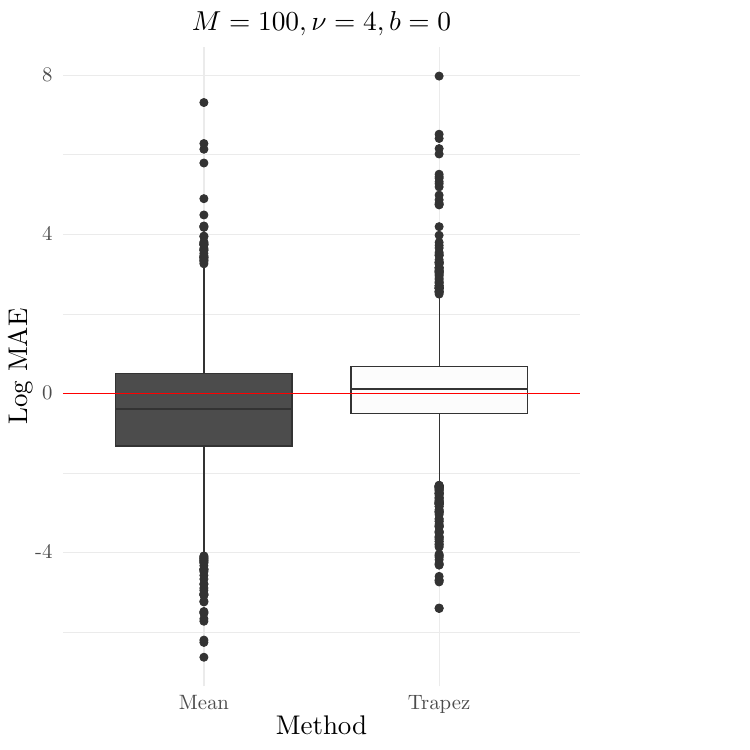}
\!\!\!\!\!\!\!\!\includegraphics[height = 0.2\textheight,width=.18\textwidth]{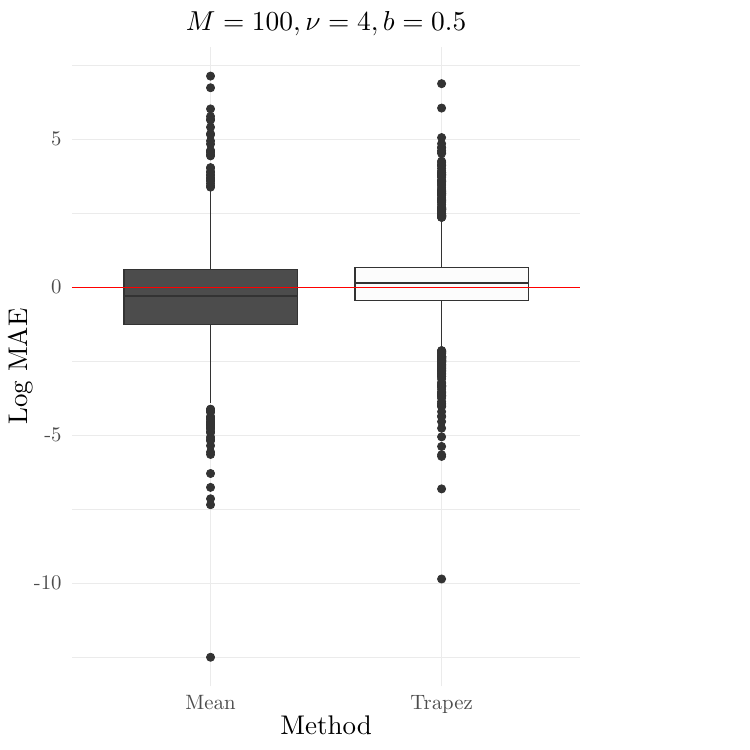}

\includegraphics[height = 0.2\textheight,width=.18\textwidth]{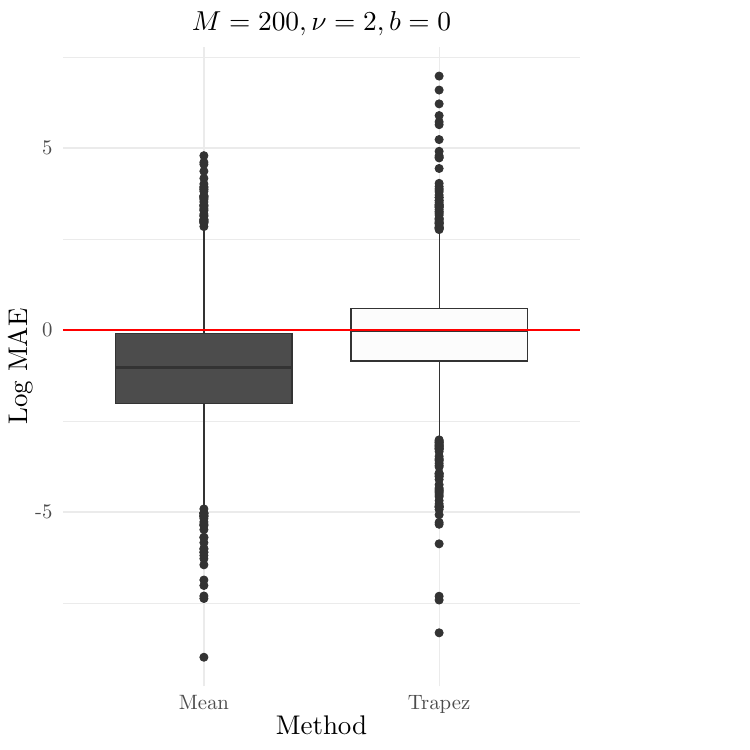}
\!\!\!\!\!\!\!\!\includegraphics[height = 0.2\textheight,width=.18\textwidth]{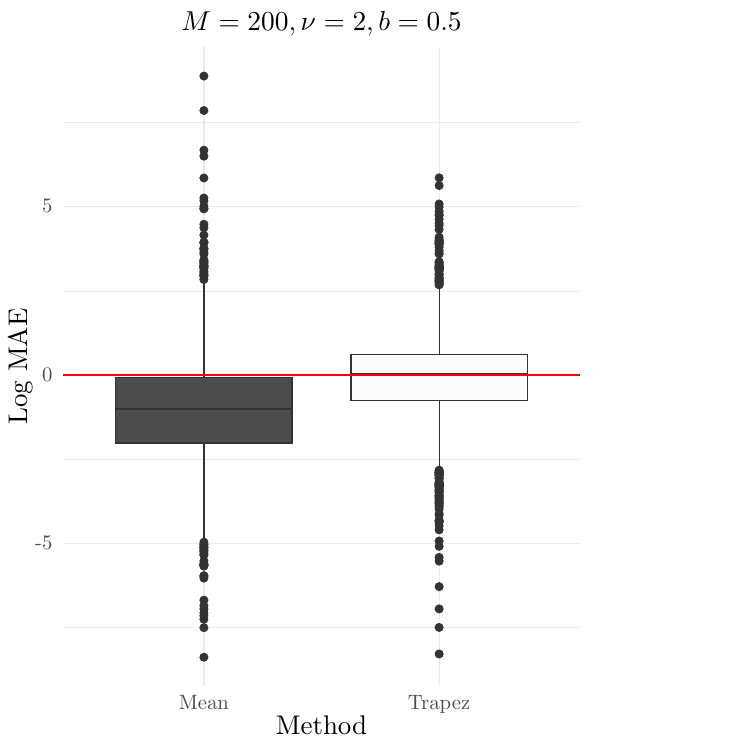}
\!\!\!\!\!\!\!\!\includegraphics[height = 0.2\textheight,width=.18\textwidth]{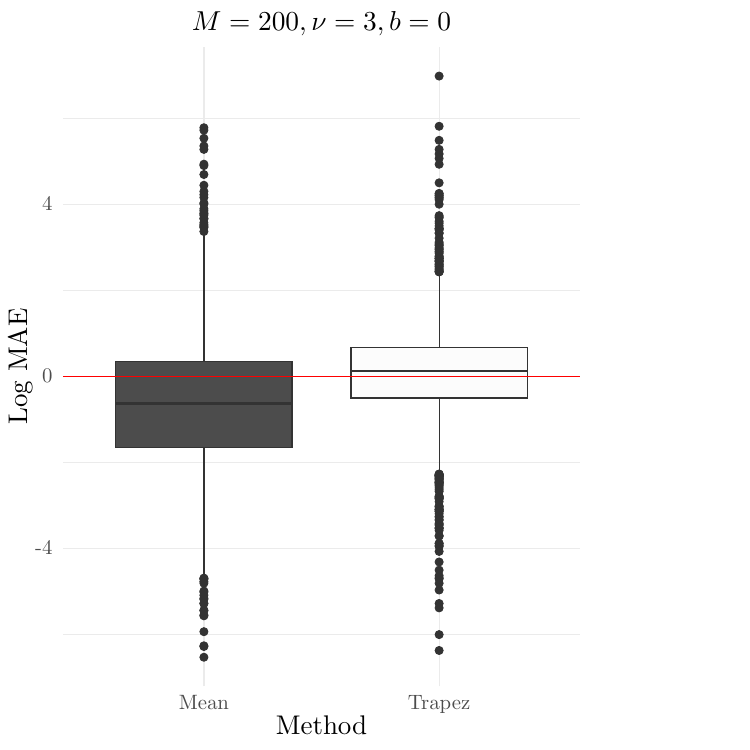}
\!\!\!\!\!\!\!\!\includegraphics[height = 0.2\textheight,width=.18\textwidth]{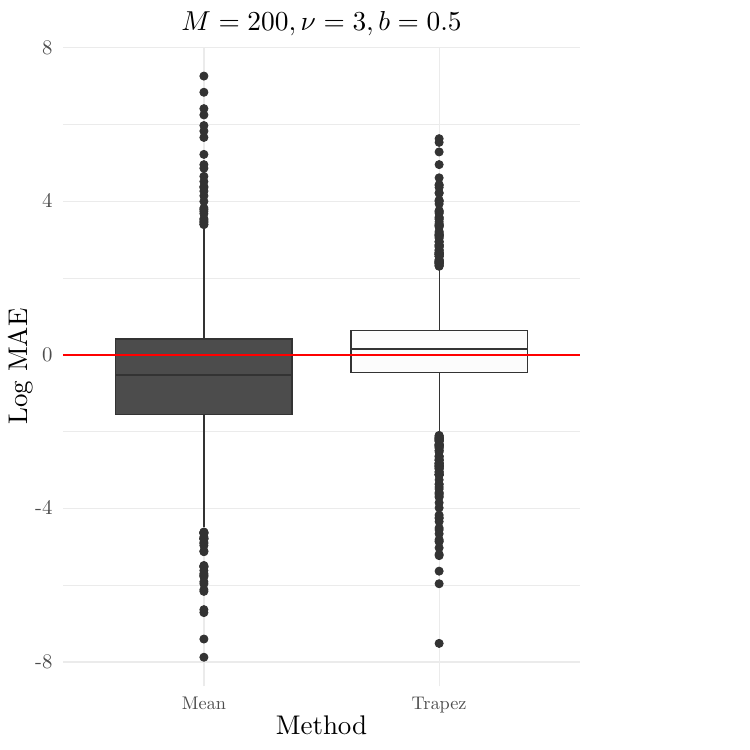}
\!\!\!\!\!\!\!\!\includegraphics[height = 0.2\textheight,width=.18\textwidth]{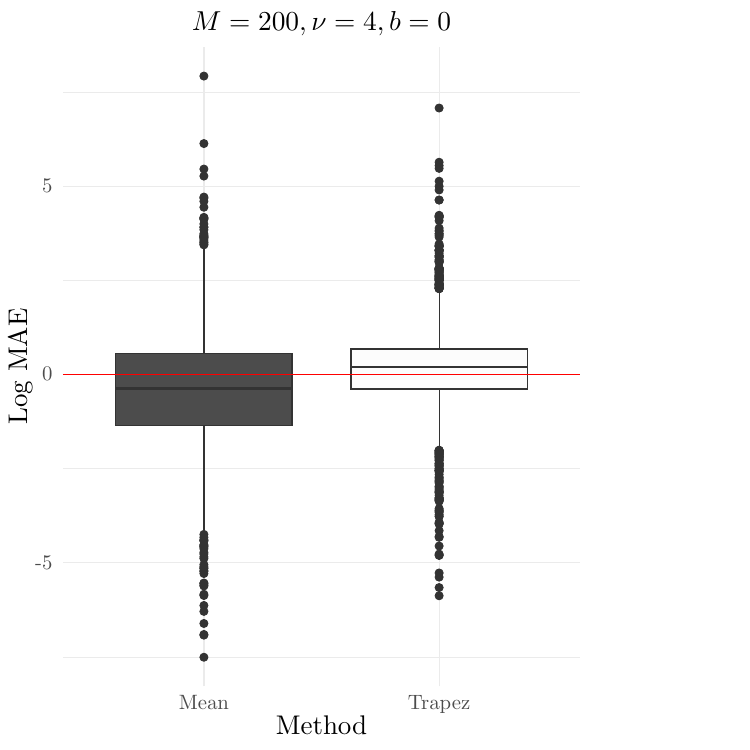}
\!\!\!\!\!\!\!\!\includegraphics[height = 0.2\textheight,width=.18\textwidth]{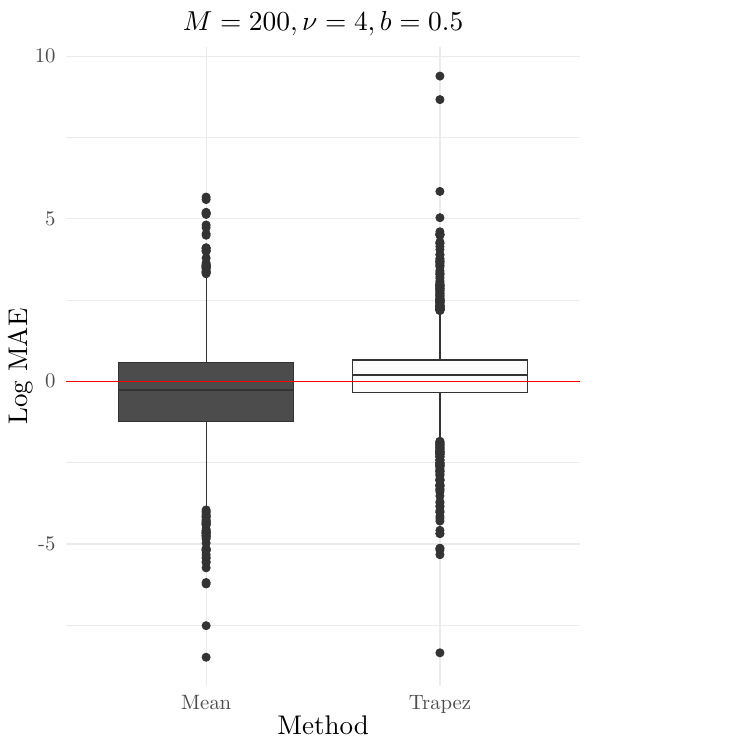}

\caption{\small Best linear prediction in linear functional regression with noisy covariate: boxplots of the log-ratios of absolute errors $\mathcal{R}(\widehat I^c(\varphi), \widehat I(\varphi)) $,  $c \in \{{\rm mean, trapez}\}$. The simulation configurations for the sample paths and the design points are like in Fig. \ref{fig:box-reg-noiseless}, the covariate noise is  $\mathcal N (0, (0.1)^2)$, the results are obtained from 2000 replications.}
\label{fig:box-reg-noisy}

\end{figure}

In the setup of noisy functional covariate as described in \eqref{noisy_regZ}, comparisons of estimates can be similarly made to both Riemann sums and sample means. We consider the same parameter settings as the noiseless case were used, with an additional noise term with constant variance $\sigma = 0.1$, and $e\sim \mathcal N (0,1)$.
Boxplots for the estimates can be seen in Figure \ref{fig:box-reg-noisy}. We see that the control neighbors method always perform better than the sample mean, and is comparable to the Riemann sums (trapezoidal rule).

In the context of constructing confidence intervals in the presence of noise, the advantage of the control neighbors approach
is given by Proposition \ref{CLT-prop}, which provides simple asymptotic  intervals. Confidence intervals are not so straightforward for Riemann sums and the sample mean. For the former, this is due to a lack of asymptotic convergence results. For the latter, although the CLT guarantees convergence of the distribution, the asymptotic variance is contaminated by the error resulting from the integral approximation, since its rate is not negligible with respect to the CLT. We therefore only consider the coverage and the lengths of the confidence intervals of the control variates method, comparing between the two variances that can be used in view of Proposition \ref{CLT-prop}, namely $s^2_M$, the conditional variance given the design, and its asymptotic expression in \eqref{CI_as}, respectively. The coverages and lengths of the confidence intervals are given in Table \ref{tab:cov-width-reg-noisy}. We denote the coverages by $p^{\rm cond}$ and $p^{\rm lim}$, while $\ell^{\rm cond}$ and $\ell^{\rm lim}$ denote the lengths of the confidence intervals. We observe that most of the coverages are close to the nominal level and the lengths are quite similar.

\begin{table}[ht]
\centering
\begin{tabular}{ccccccc}
  \hline
$M$ & $\nu$ & $b$ & $p^{\rm cond}$ & $p^{\rm lim}$ & $\ell^{\rm cond}$ & $ \ell^{\rm lim}$ \\ 
  \hline
 50 &   2 & 0.0 & 0.90 & 0.90 & 0.36 & 0.36 \\ 
   50 &   2 & 0.5 & 0.91 & 0.91 & 0.34 & 0.34 \\ 
   50 &   3 & 0.0 & 0.95 & 0.95 & 0.35 & 0.35 \\ 
   50 &   3 & 0.5 & 0.96 & 0.96 & 0.33 & 0.33 \\ 
   50 &   4 & 0.0 & 0.95 & 0.95 & 0.35 & 0.35 \\ 
   50 &   4 & 0.5 & 0.96 & 0.96 & 0.33 & 0.33 \\ \hline
  100 &   2 & 0.0 & 0.93 & 0.93 & 0.26 & 0.26 \\ 
  100 &   2 & 0.5 & 0.93 & 0.93 & 0.24 & 0.24 \\ 
  100 &   3 & 0.0 & 0.94 & 0.94 & 0.25 & 0.25 \\ 
  100 &   3 & 0.5 & 0.94 & 0.94 & 0.23 & 0.24 \\ 
  100 &   4 & 0.0 & 0.94 & 0.94 & 0.25 & 0.25 \\ \hline
  100 &   4 & 0.5 & 0.94 & 0.94 & 0.23 & 0.24 \\ 
  200 &   2 & 0.0 & 0.94 & 0.94 & 0.182 & 0.182 \\ 
  200 &   2 & 0.5 & 0.94 & 0.94 & 0.170 & 0.17 \\ 
  200 &   3 & 0.0 & 0.95 & 0.95 & 0.176 & 0.177 \\ 
  200 &   3 & 0.5 & 0.95 & 0.95 & 0.167 & 0.167 \\ 
  200 &   4 & 0.0 & 0.94 & 0.94 & 0.175 & 0.176 \\ 
  200 &   4 & 0.5 & 0.95 & 0.95 & 0.167 & 0.167 \\ 
   \hline
\end{tabular}
\caption{\small Inference in linear functional regression:
coverage and length of confidence intervals (CI) for the mean value of the response given the noisy covariate observations. 
CI based on the CLT for the control neighbor estimates, using the conditional variance $s_M^2$ or its limit \eqref{CI_as}.}
\label{tab:cov-width-reg-noisy}
\end{table}


\subsection{Scores approximation}

We focus on the bivariate case with $\mathcal{T} = [0,1]^2$. Let $X$ be a generic sample path, a surface in this case. 
The 2-dimensional random design points $T_m$, $1\leq m \leq M$, are obtained as copies of the bivariate vector $T$ which admits the density $f_{T}$. Recall that the integrand for the $j$-th score $X$ is given by
\begin{equation}
\varphi_j(t) = \frac{\left\{X( t) - \mu(t \right\} \psi_j(t)}{f_{T}(t)}, \qquad t\in\mathcal T. 
\end{equation}
The basis  functions $\psi_j$ and the density $f_T$ are assumed to be given, while the values of $X$ at the design points are assumed noiseless. 

The unbiased control neighbors estimate of the integral of $\varphi_j$ over $\mathcal T$ requires one to build the Voronoi diagram $M$ times, a computationally heavy task when $d>1$. Following \cite{leluc2024speeding}, the unbiased leave-one-out control neighbors estimate can be replaced by its computationally efficient counterpart, given by
\begin{equation}\label{eq:cnn-biased}
\widehat I^{{\rm (NN)}}(\varphi) = \frac{1}{M}\sum_{m=1}^M \varphi( T_m) - \frac{1}{M}\sum_{m=1}^M \widetilde \varphi^{(m)} (T_m) + \sum_{m=1}^M \varphi(T_m) V_{M, m},
\end{equation}
where $\widetilde \varphi^{(m)} $ is the LOO-1NN estimate  and $V_{M,m}$ is the volume of the Voronoi cell of the design point $T_m$, both  based on the full sample $ T_1, \dots, T_M$. See also the Appendix \ref{CV_1NN}. 
Although $\widehat I^{{\rm (NN)}}(\varphi)$ is biased, the root mean squared distance between $\widehat I^{{\rm (NN)}}(\varphi)$ and $\widehat I (\varphi)$ is of order $O_{\mathbb{P}}(M^{-1/2-\beta/d})$, which means that the two estimates have the same fast
convergence rate. However,\eqref{eq:cnn-biased} is much easier to compute, since the Voronoi diagram only needs to be computed once. In view of computational efficiency, we adopt the version in \eqref{eq:cnn-biased} for our simulations, and recommend the version $\widehat I^{{\rm (NN)}}(\varphi)$ whenever $d > 1$. 

The design points were simulated as 2-dimensional random vectors with independent uniform components. 
Surfaces were simulated using a truncated version of the multivariate Kosambi-Karhunen-Loève (KKL) decomposition of the bivariate mean centered Wiener sheet; see \cite{Deheuvals2006}. With $\{\omega_{k_1, k_2}: k_1, k_2 \geq 1\}$ denoting an array of i.i.d standard Gaussian random variables, we define 
\begin{equation}\label{eq:KL-2D}
X(t) = \sum_{k_1=1}^{K_1} \sum_{k_2=1}^{K_2} \omega_{k_1, k_2} \frac{\sqrt{2}\cos(k_1 \pi t^{(1)})}{(k_1\pi)^{\gamma_1}}\frac{\sqrt{2}\cos(k_2 \pi t^{(2)}) }{(k_2 \pi)^{\gamma_2}}, \qquad \forall t = (t^{(1)},t^{(2)})\in [0,1]^2 , 
\end{equation}
and we use this representation to simulate a surface on a random grid of points. The process $X$ in \eqref{eq:KL-2D} becomes the bivariate Wiener sheet if  $\gamma_1=\gamma_2=1$ and $K_1, K_2 = \infty$. The terms $(k_1 \pi)^{-\gamma_1}$ and $(k_2 \pi)^{-\gamma_2}$ represents the square root of the $k_1-$th and $k_2-$th eigenvalue, respectively. Here, we allow the rate of decay of eigenvalues to vary, allowing us to adjust the smoothness of the integrand. 

The numbers of basis functions $K_1$, $K_2$ were set to $K_1 = K_2 = 12$, since a small number of basis functions already captures most of the explained variance. This can be seen in Table \ref{tab:explained-var-2d}. The scores are given by
\begin{equation}
\xi_{k_1, k_2} = \frac{\omega_{k_1, k_2}}{(k_1\pi)^{\gamma_1} (k_2\pi)^{\gamma_2}}, 
\end{equation}
and we focus our attention on the recovery of the first three scores on the diagonal $\{\xi_{1,1}, \xi_{2,2}, \xi_{3,3}\}$.
Comparisons were made to the sample mean estimator for the configurations consisting of all combinations of the parameters  $\gamma_1=\gamma_2 \in \{1, 1.5, 2\}$, $M \in \{50, 100, 200\}$. Prediction intervals were similarly constructed according to Section \ref{sec:noiseless-inf}, where the Hölder exponent was set to $\beta = \min\{\gamma_1, \gamma_2\} - 1/2$, and 1000 subsamples were used.

\begin{figure}[!ht]
\centering
\includegraphics[height = 0.2\textheight,width=.3\textwidth]{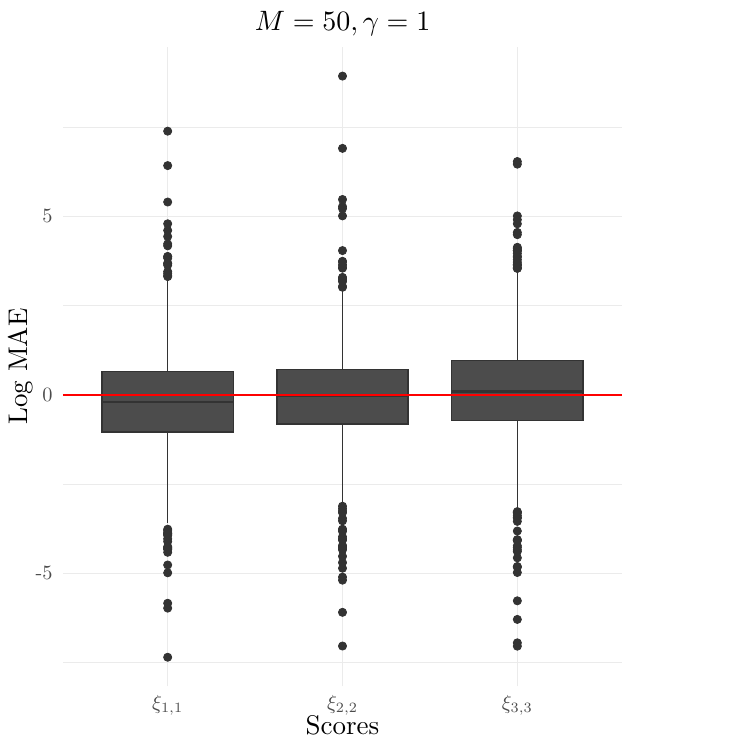}
\includegraphics[height = 0.2\textheight,width=.3\textwidth]{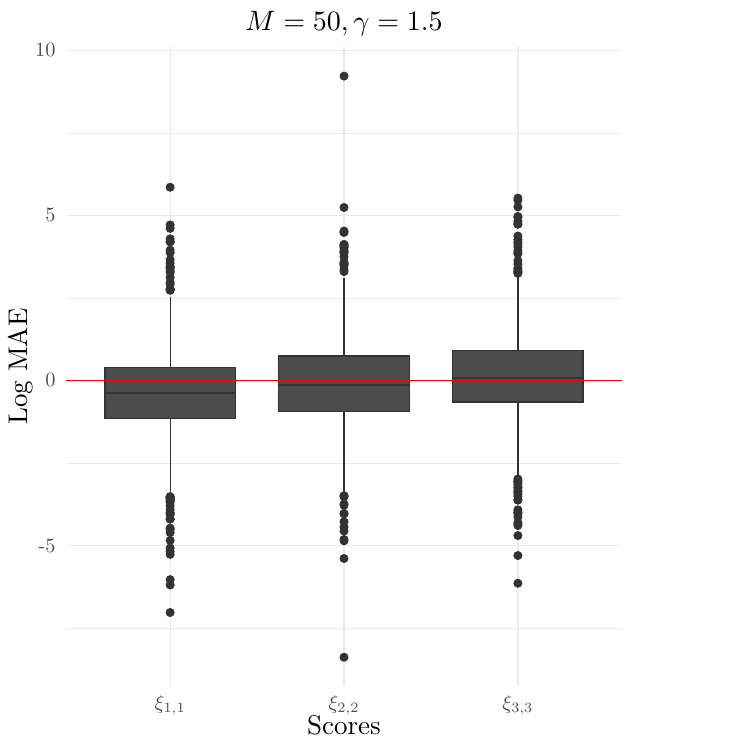}
\includegraphics[height = 0.2\textheight,width=.3\textwidth]{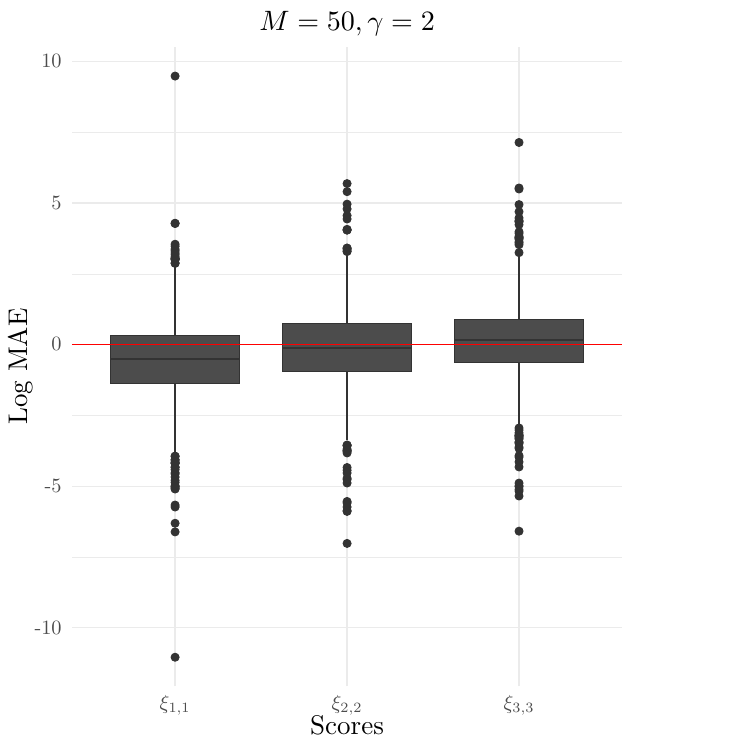}

\includegraphics[height = 0.2\textheight,width=.3\textwidth]{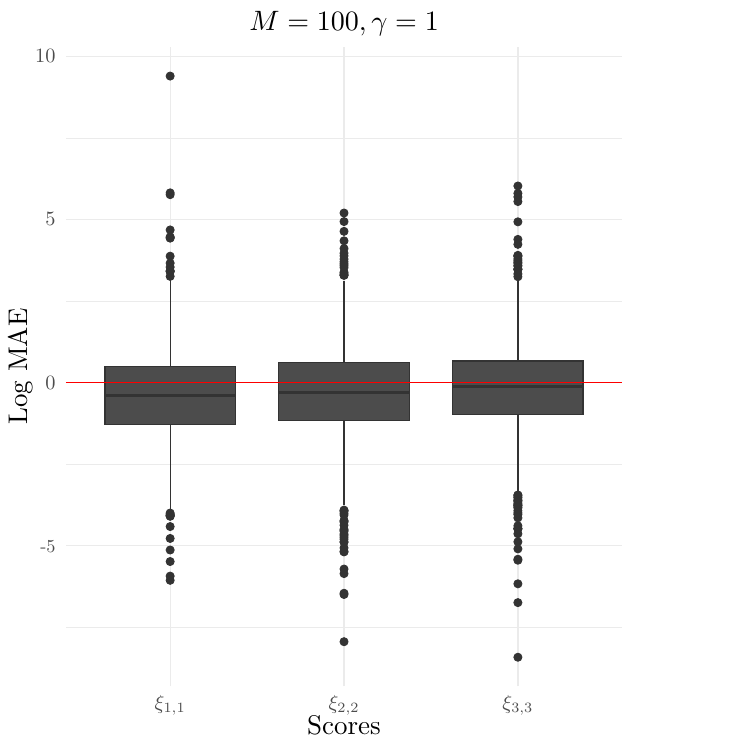}
\includegraphics[height = 0.2\textheight,width=.3\textwidth]{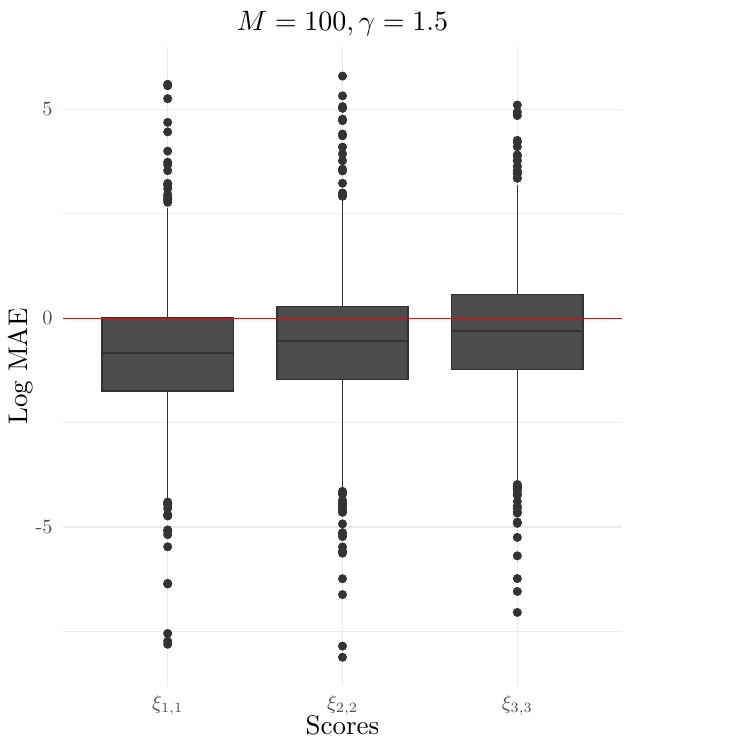}
\includegraphics[height = 0.2\textheight,width=.3\textwidth]{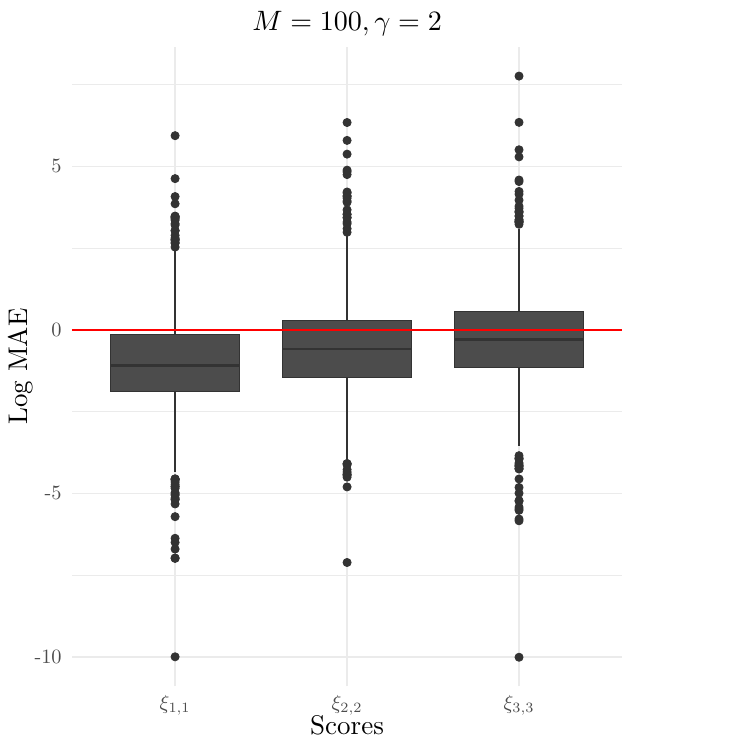}

\includegraphics[height = 0.2\textheight,width=.3\textwidth]{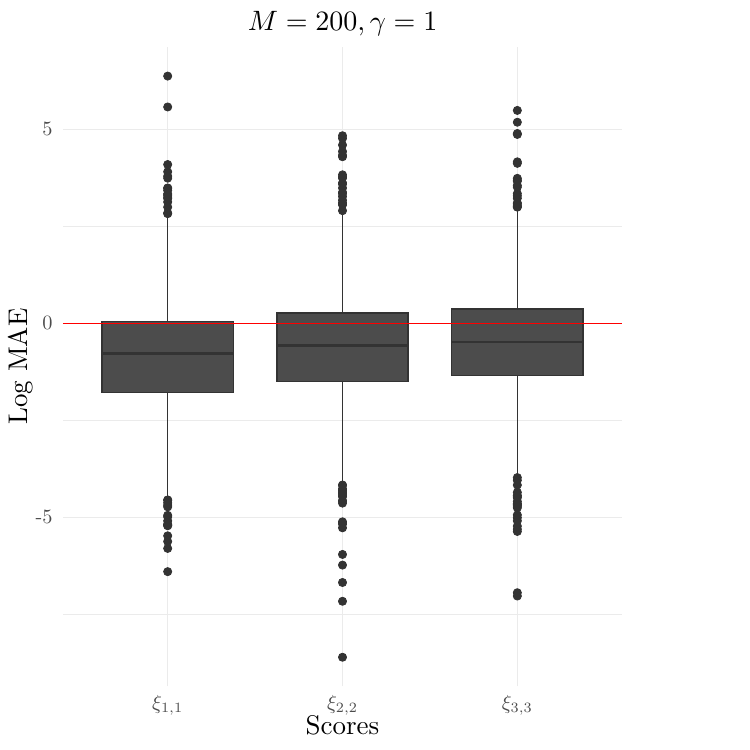}
\includegraphics[height = 0.2\textheight,width=.3\textwidth]{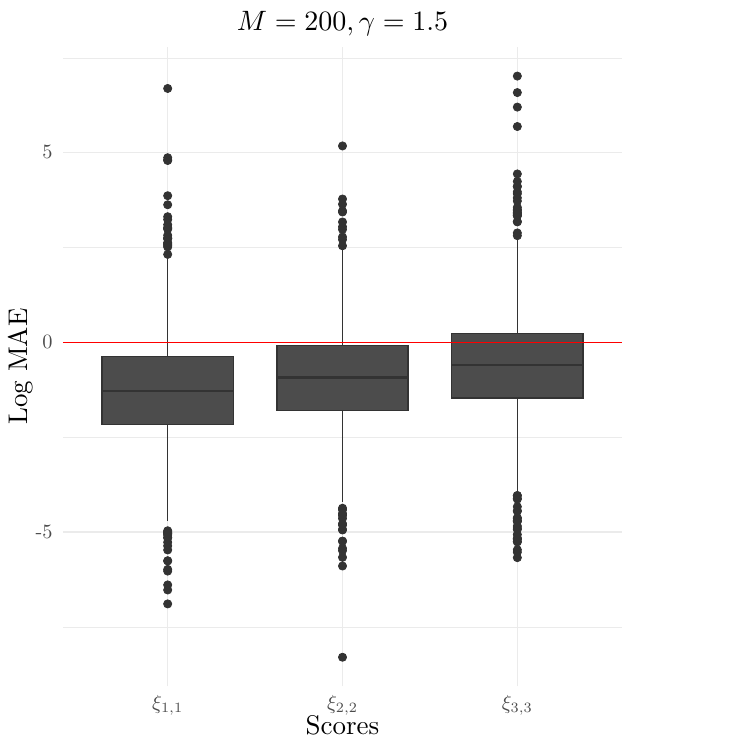}
\includegraphics[height = 0.2\textheight,width=.3\textwidth]{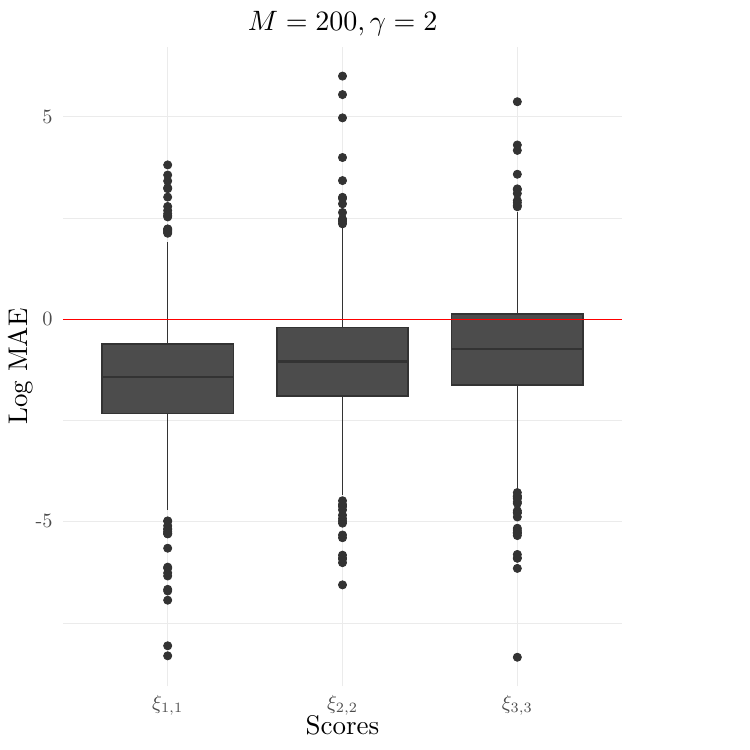}

\caption{\small Boxplots showing the log-ratios of absolute errors of the scores estimates for the control neighbors  and the sample mean approaches, the case of bivariate random functions generated according to \eqref{eq:KL-2D}, with $K_1=K_2=12$. Results below the zero level indicate a better performance for the control neighbors estimates. 
Different configurations for regularity (with $\gamma_1=\gamma_2$), and bivariate design points sample sizes $M$.}
\label{fig:box-scores}

\end{figure}

Boxplots for the estimates can be seen in Figure \ref{fig:box-scores}. We see that the errors are always at least as good as the sample mean, and much better in certain configurations. 
The coverage and the relative lengths of the prediction intervals of nominal level $1 - \delta = 0.95$ can be seen in Table \ref{tab:scores-cov-width}. We see that the control neighbors approach generally yields accurate coverage and  better lengths.

\begin{table}[]
\centering
\begin{tabular}{c|c|c|c|}
  \backslashbox{\!\!$K_1 \!=\! K_2$}{$\gamma_1\!=\!\gamma_2$}     & 1    & 1.5   & 2    \\
   \hline
  3 & 68.8 & 93.5  & 98.6 \\
\hline
    4 & 75.4 & 96.0    & 99.4 \\
   \hline
    5 & 79.7 & 97.3  & 99.6
\end{tabular}
\caption{\small Bivariate random functions: the different levels of explained variance according to the number of basis functions $K_1=K_2$ in the representation \eqref{eq:KL-2D} for different levels of smoothness $ \gamma_1=\gamma_2$.}
\label{tab:explained-var-2d}
\end{table}

\begin{table}[ht]\small 
\centering
\begin{tabular}{cccccc}  
  \hline
$M$ & $\gamma_1=\gamma_2$ & $\xi$ &  $p^{{\rm (NN)}}$ & $p^{\rm (m)}$ & $(\ell^{{\rm (NN)}} -\ell^{\rm (m)}) / \ell^{\rm (m)}$ \\ 
  \hline
    &     & $\xi_{1,1} $    & 93.2   & 79.6   & 0.1296 \\ 
      & 1.0   & $\xi_{2,2} $    & 91.6   & 80.9   & 0.1968 \\ 
      &     & $\xi_{3,3}$    & 91.4   & 80.8   & 0.3162 \\ \hline
      &     & $\xi_{1,1}$    & 91.1   & 81.5   & -0.1549 \\ 
     50 & 1.5   & $\xi_{2,2}$    & 86.8   & 77.5   & -0.0326 \\ 
      &    & $\xi_{3,3}$    & 85.5   & 81.3   & 0.1142 \\ \hline
      &     & $\xi_{1,1}$    & 93.2   & 81.2   & -0.2105 \\ 
      & 2.0   & $\xi_{2,2}$    & 88.6   & 81.9   & -0.0618 \\ 
      &     & $\xi_{3,3}$    & 84.3   & 83.5   & 0.1100 \\ \hline\hline
     &    & $\xi_{1,1}$    & 96.2   & 83.2   & -0.0871 \\ 
     & 1.0   & $\xi_{2,2}$    & 96.1   & 85.4   & 0.0206 \\ 
     &    & $\xi_{3,3}$    & 94.0   & 80.8   & 0.1506 \\ \hline
     &     & $\xi_{1,1}$    & 95.4   & 84.7   & -0.4027 \\ 
    100 & 1.5   & $\xi_{2,2}$    & 91.6   & 81.5   & -0.2414 \\ 
     &     & $\xi_{3,3}$    & 91.1   & 81.3   & -0.0616 \\ \hline
     &     & $\xi_{1,1}$    & 96.3   & 84.3   & -0.4633 \\ 
     & 2.0   & $\xi_{2,2}$    & 94.2   & 82.9   & -0.2788 \\ 
     &     & $\xi_{3,3}$    & 91.5   & 81.1   & -0.0679 \\ \hline\hline
     &    & $\xi_{1,1}$    & 97.8   & 82.9   & -0.2804 \\ 
     & 1.0   & $\xi_{2,2}$    & 97.6   & 85.7   & -0.1970 \\ 
     &     & $\xi_{3,3}$    & 97.1   & 83.9   & -0.0693 \\ \hline
     &     & $\xi_{1,1}$    & 97.0   & 81.3   & -0.5794 \\ 
    200 & 1.5   & $\xi_{2,2}$    & 95.0   & 82.1   & -0.4595 \\ 
     &     & $\xi_{3,3}$    & 95.0   & 84.5   & -0.2793 \\ \hline
     &    & $\xi_{1,1}$    & 97.4   & 82.3   & -0.6476 \\ 
     & 2.0   & $\xi_{2,2}$    & 96.6   & 83.9   & -0.4963 \\ 
     &     & $\xi_{3,3}$    & 95.5   & 83.7   & -0.2957 \\ 
   \hline
\end{tabular}
\caption{\small  Coverage $p$ and lengths $\ell$ of prediction intervals for the scores of a bivariate random function generated according to \eqref{eq:KL-2D}, with $K_1=K_2=12$ and different values $\gamma_1=\gamma_2$. Comparison of the  control neighbors {\rm (NN)} and the sample mean {\rm (m)} approaches for different sample sizes $M$ of  design points in $[0,1]^2$. }
\label{tab:scores-cov-width}
\end{table}

\section{Data Application}\label{sec:data-appli}
In this section, our methodology is applied on real sports data. The data set\footnote{Available at https://github.com/ArthurLeroy/MagmaClustR/blob/master/data/swimmers.rda} contains 3456 performance curves of male and female French athletes between the ages of 10 and 20 for the 100m freestyle event. An important task in the analysis of sports data is clustering, where the aim is to distinguish the best athletes from the rest. In the FDA framework, clustering is commonly performed on the fPCA scores, which is our goal. Comparisons to the trapezoidal rule and Riemann sums will be made.

\begin{figure}
\centering
\includegraphics[scale=0.15]{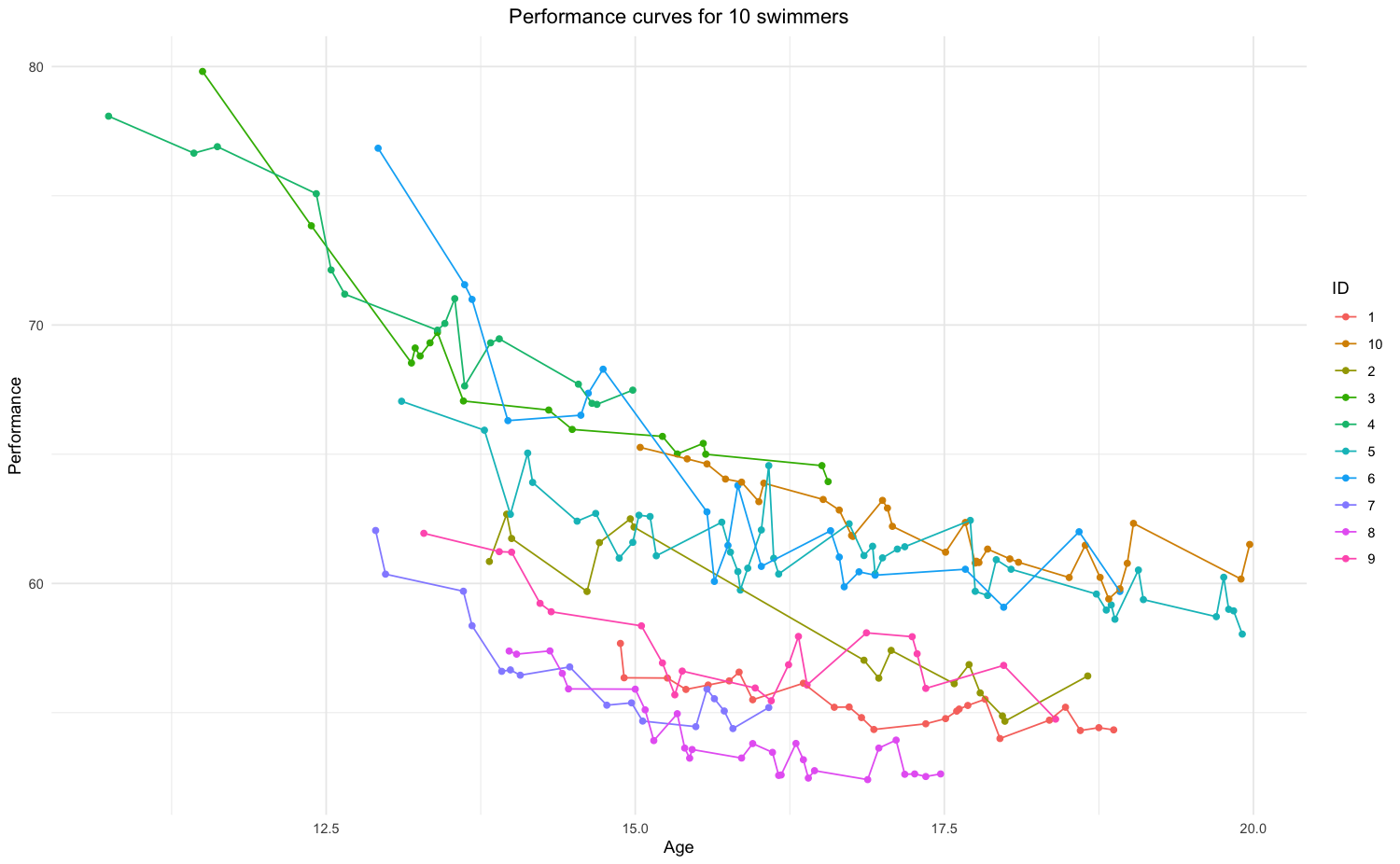}
\caption{\small Performance curves of the first 10 swimmers.}
\label{fig:swim-10}
\end{figure}

Observations of athlete performance are usually considered to be noiseless, since they are recorded with high precision sensors. The original domain was $\mathcal{T} = [10, 20]$, with the design points representing the random age at which the athletes compete. Ages were normalized to be in $\mathcal{T} = [0, 1]$ by subtracting the minimum age and dividing by the range. Plots of the first 10 swimmers before rescaling can be seen in Figure \ref{fig:swim-10}. Many of the curves are sparse, with only a few observed points per curve.


Recall the score equation \eqref{scores-eq}. In order to compute the scores in practice, the auxiliary quantities $\mu$, $\psi_j$ and $f_T$ need to be estimated from the data. Since we are focused on the integration aspect, and not the estimation of auxiliary quantities, the same methods will be used for comparisons against other integration methods. We now briefly describe the estimation procedures of the auxiliary quantities. 

\begin{remark}
In FDA, pooling the observation points across subjects gives the practitioner access to $\overline M = \sum_{i=1}^n M_i$ points for several quantities of interest. This is true for the auxiliary quantities mentioned above, so the rate of convergence for estimating them is expected to be negligible with respect to the rate of integral approximation along one curve. 
\end{remark}

The density $f_T$ is estimated using series expansions with thresholding. Let $c_{TH}, c_{k0}$ and $c_{k1}$ be constants, and $\{\Phi_k \}_{k=1}^K$ be the orthonormal cosine basis given by $\Phi_0 = 1$ and $\Phi_k(t) = \sqrt{2}\cos(\pi k t), \forall k \geq 1$. Denote $\mathbbm{1}\{.\}$ to be the indicator function. The thresholding estimator (\cite{efromovich2018}) is given by 
\begin{equation}
\widehat f_T(t) = \sum_{k=0}^{\widehat{K}} \widehat \theta_k \mathbbm{1}{\left\{\widehat \theta_k^2 > c_{TH} \widehat v_{k} \right\}} \Phi_k(t), 
\end{equation}
where $\widehat \theta_k = n^{-1}\sum_{i=1}^n M_i^{-1} \sum_{m=1}^{M_i} \Phi_k(T_{i,m})$ is the pooled sample mean, and $\widehat v_k$ is the sample variance estimate of $\widehat \theta_k$. The empirical cutoff $\widehat K$ is an integer selected by the rule
\begin{equation}
\widehat K = \arg\min_{0 \leq K \leq c_{K0} + c_{K1} \log(\overline M)} \left\{\sum_{k=0}^K 2\widehat v_{k} - \widehat \theta_k^2 \right\}. 
\end{equation}
Following \cite{efromovich2018}, the thresholding constants were chosen to be $c_{k0} = 3, c_{k1} = 0.8$ and $c_{TH} = 0.4$. 

\begin{figure}[ht!]
    \centering
        \includegraphics[width=0.3\textwidth, height = 5cm]{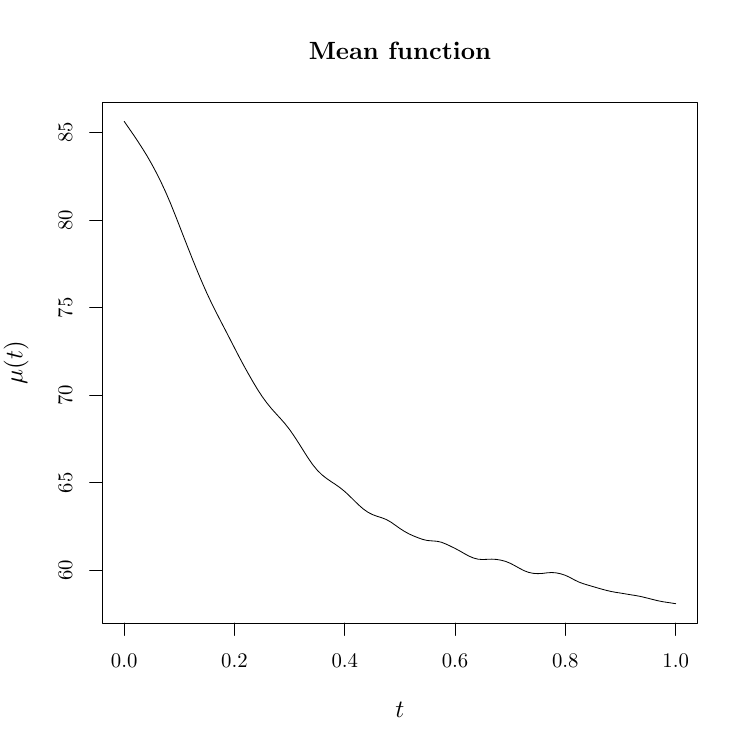}        \includegraphics[width=0.3\textwidth, height = 5cm]{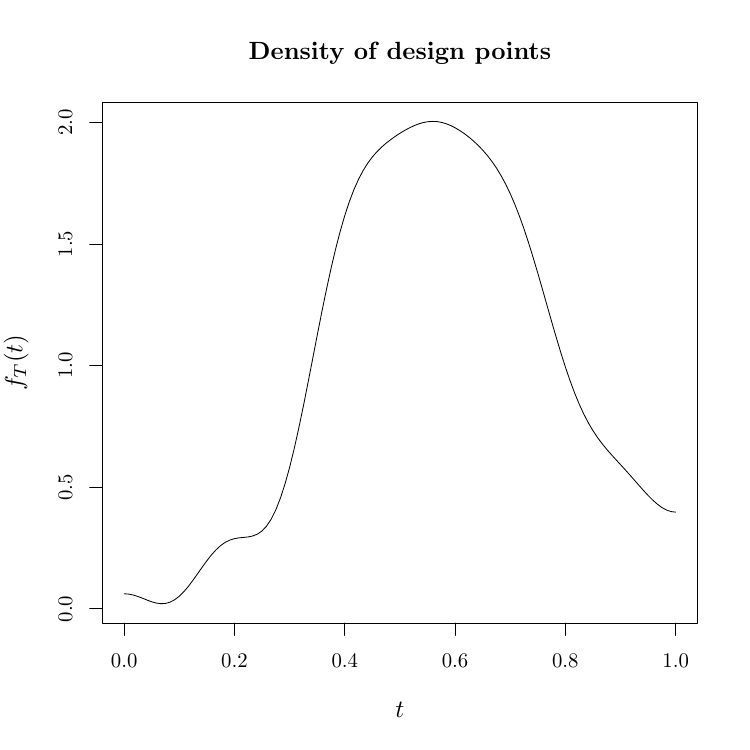}
        \includegraphics[width=0.3\textwidth, height = 5cm]{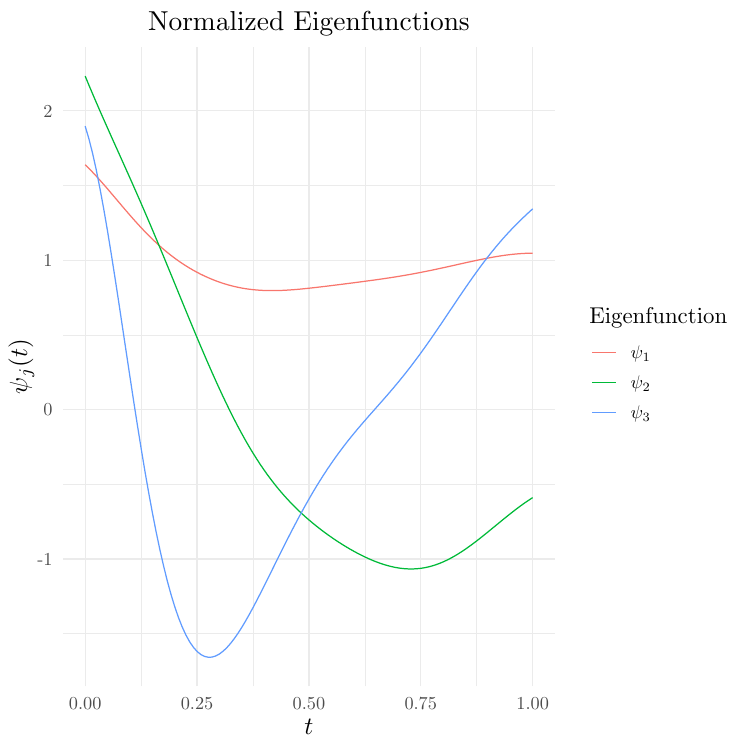}
        \caption{Estimated mean function $\mu$, density $f_T$, and the normalized eigenfunctions $\{\psi_j\}_{j=1}^J$ respectively. $J=3$ were selected by fraction of explained variance. Eigenfunctions were normalized to the same sign.}
\label{fig:perf-aux}
\end{figure}

The mean function is estimated by applying a smoothing splines estimator on the pooled data points, with the smoothing parameter $\lambda$ chosen by generalized cross-validation. See \cite{caiyuan2012reg}. The eigenfunctions were estimated by applying a local polynomial estimator on the pooled data points; see \cite{Yao2005}. Due to computational difficulties, the default bandwidth of 0.1 was used. The number of eigenfunctions were selected by the fraction of explained variance (FEV), with the threshold set to 0.95. The equally spaced estimation grid for all the auxiliary quantities were chosen to have a resolution of $1/120$, corresponding to one month over 10 years. Plots of the auxiliary quantities can be seen in Figure \ref{fig:perf-aux}.

\begin{remark}
Although the scores can also be estimated using the fPCA method of \cite{Yao2005} by means of conditional expectation, it is tailored for the noisy setup, which makes it unsuitable for the analysis of swimmers' performance curves. Moreover, the selection of a data-driven bandwidth in \cite{Yao2005} remains a tricky issue, due to computationally difficulties quickly encountered with cross-validation after pooling the observation points. 
\end{remark}

Linear interpolation was performed from the estimation grid to the observed points for the auxiliary quantities to construct $\varphi_j(T_m^{(i)})$. The score estimates were scaled by the empirical standard deviation, corresponding to the square root of eigenvalues. The hierarchical clustering algorithm with average linkage and Euclidean distance was applied to the random vector of scores, with $L = 2$ clusters selected. Summary statistics of cluster separation can be found in Table \ref{tab:clust-n-appli}. 

\begin{figure}[ht!]
    \centering
        \includegraphics[width=0.3\textwidth, height = 5cm]{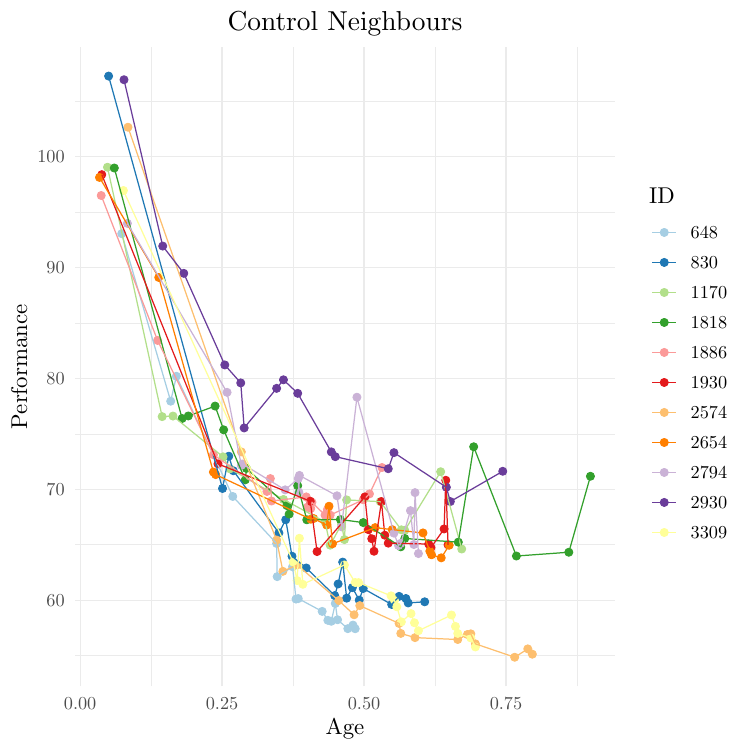}
        \includegraphics[width=0.3\textwidth, height = 5cm]{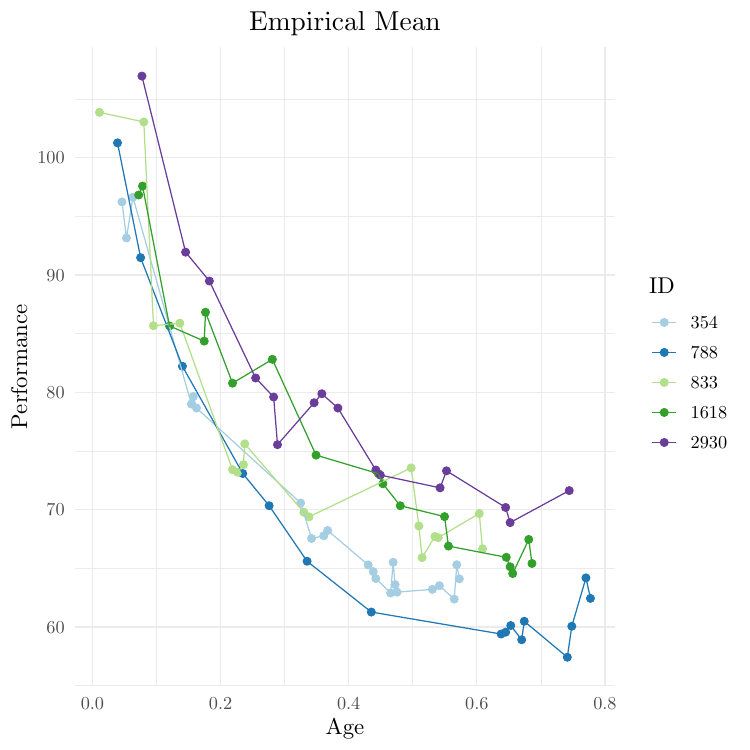}
        \includegraphics[width=0.3\textwidth, height = 5cm]{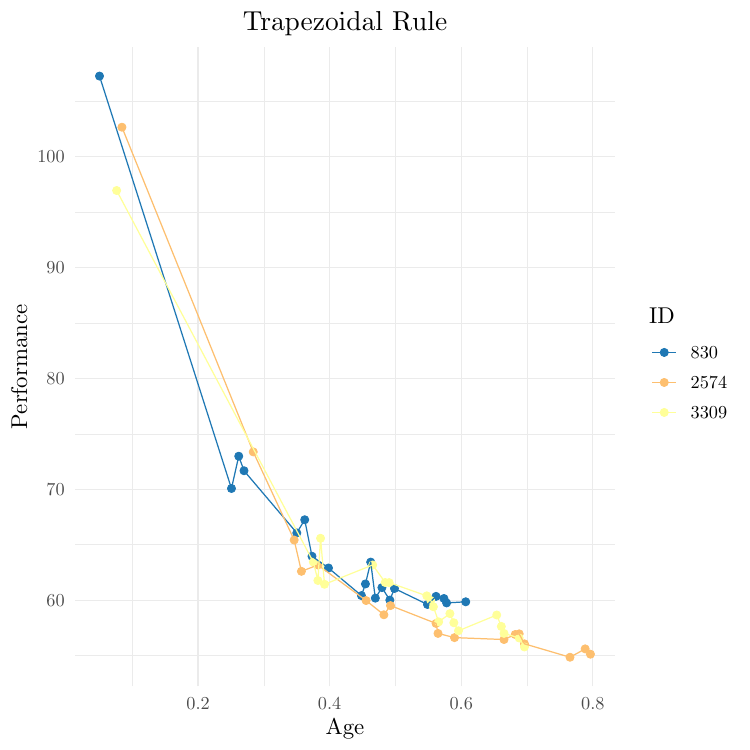}
        \caption{\small Performance curves selected into the smaller cluster $L_2$ for the different integration methods. }
        \label{fig:perf-curve-clust}
\end{figure}

Different cluster sizes for $L_1$ and $L_2$ were observed for the different integral approximation methods, with the control neighbors method selecting the largest number of individuals into the smaller group. The $L_2$ cluster can be interpreted to be the group of athletes with the largest improvement in performance, as seen from the range $R_2$. Plots of the athletes selected in $L_2$ is provided in Figure \ref{fig:perf-curve-clust} for the different methods. 

\begin{table}[]
\centering
\begin{tabular}{l|l|l|l|l}
   & $\widehat \xi^{(NN)}$  & $\widehat \xi^{(m)} $  & $\widehat \xi^{(trapez)}$ \\
   \hline
$L_1$ & 3445 & 3451 & 3453  \\
$L_2$ & 11   & 5    & 3    \\
$R_1$ & 11.77 & 11.81 & 11.82 \\
$R_2$ & 36.96 & 37.41 & 45.54 
\end{tabular}
\caption{\small The number of individuals partitioned into the clusters $L_1$ and $L_2$ for the different methods. $R_1$ and $R_2$ denotes the range of performance times of individuals in clusters $L_1$ and $L_2$ respectively.  Range is calculated by the difference of the maximum and minimum performance times.}
\label{tab:clust-n-appli}
\end{table}


\section{Acknowledgements}
The authors gratefully acknowledge funding from the French National Research Agency within the framework of the France 2030 programme for EUR DIGISPORT (ANR-18-EURE-0022) project. The authors thank François Portier for helpful discussions related to the paper.

\clearpage

\newpage

\section{Appendix}\label{sec:appendix}

\subsection{Control variates definitions and main properties}\label{CV_1NN}

We here recall the main definitions and properties related to the control variates with nearest neighbor, as presented by \cite{leluc2024speeding}. With   $ \widehat N^{(m)}(t)$ defined in \eqref{NN_t}, a simplified notation for the LOO-NN $\widehat N_M^{(m)}(t)$, the leave-one-out Voronoi cells are given by
\begin{equation}
	\forall \ell \in \{1, \dots, M\} \textbackslash \{ m \}, \qquad S^{(m)}_{\ell} = S^{(m)}_{M, \ell} = \left\{t \in \mathcal{T}: \widehat N^{(m)}(t) = T_{\ell} \right\}.
\end{equation}

\begin{definition} \label{def_weights}
	Let $M$ be a given positive integer, and $T_1,\ldots,T_M\in\mathcal T \subset \mathbb R^d$.
\begin{enumerate}
\item \emph{(Degree)}	For all $\ell = 1,\dots, M$, the degree $\widehat d_{\ell  }  $ represents the number of times $T_\ell$ is a nearest neighbor of a point $T_m$ for all $m\neq \ell $. Formally, 
$ \widehat d_{\ell }  = \widehat d_{M,\ell}  = \sum_{m:m\neq \ell}\mathbbm{1}_{S^{(m)}_{\ell}}(T_m)$.
\item \emph{(Cumulative Voronoi Volume)}	The cumulative volume is given by
$ \widehat c_{\ell} = \widehat c_{M,\ell} = \sum_{m:m\neq \ell} V^{(m)}_{\ell}$, where $V^{(m)}_{\ell}= V^{(m)}_{M,\ell} = \mathbb P (T \in S^{(m)}_{\ell})$.
\end{enumerate}
\end{definition}

\begin{proposition}[Lemma 2 and Proposition 1, \cite{leluc2024speeding}]\label{prop_c_d}	
	Assume that $T_1,\ldots , T_M$ are random copies of $T\in\mathcal T$, independent of $M$. It holds that  
	$\mathbb E _M [ \widehat d_{m}  ]=\mathbb E_M \left[ \widehat c_{m} \right]=1$, and 
	$$
	\sum_{m=1}^M \widehat{d}_{m} \varphi (T_m) = \sum_{m=1}^M   \widetilde \varphi^{(m)} (T_m),\qquad  	\sum_{m=1}^M \widehat{c}_{m} \varphi (T_m) = \sum_{m=1}^M   I \left(\widetilde \varphi^{(m)}\right),
	$$
	where $\widetilde \varphi^{(m)}(t) = \varphi(\widehat N^{(m)}(t))$. In particular, 
	$\mathbb E_M \left[ \widehat{I} (\varphi) \right] = I(\varphi)$ with $\widehat{I} (\varphi)$ defined in \eqref{control-neighbors-estim}, and 
	$$
	\widehat I(\varphi) = \sum_{m=1}^M w_{M,m} \varphi(T_m) \qquad  \text{ with } \qquad w_{M,m} = (1+ \widehat{c}_{m} - \widehat{d}_{m})/M.
	$$
\end{proposition}

It is worth noting that there is a version of $	\widehat I(\varphi) $ which requires less numerical effort for $d > 1$ at the cost of a negligible bias. More precisely, \cite[Proposition 1]{leluc2024speeding} also consider
\begin{equation}\label{I_phi_NN}
\widehat I^{\rm (NN)}(\varphi) = \sum_{m=1}^M w^{\rm (NN)}_{M,m} \varphi(T_m) \qquad  \text{ with } \qquad w^{\rm (NN)}_{M,m} = (1+ M V_{M,m} - \widehat{d}_{m})/M,
\end{equation}
where $V_{M,m} $ is the Voronoi volume  $\mathbb P (T \in S_{M,m})$ and $S_{M,m}$ is the standard Voronoi cell of $T_m$. It can be shown that $\widehat I^{\rm (NN)}(\varphi) $ has the same rate of convergence as $\widehat I(\varphi) $.
 
 \medskip

 \subsection{Proofs}\label{proofs_app}

 \begin{proof}[Proof of Proposition \ref{CLT-prop}]
Let us simplify notation and write $w_m$ (resp. $w^{\rm (NN)}_{m}$) (resp. $V_{m}$) instead of $w_{M,m}$ (resp. $w^{\rm (NN)}_{M,m}$) (resp. $V_{M,m}$). Thus, in view of \eqref{eq:control-neighbors-noisy},  $\widehat \Sigma := \sum_{m=1}^M w_m \sigma_\eta(T_m)\eta_m$ and $R := \widehat I(\varphi) - I(\varphi)$. 
 Since, by Proposition \ref{th1_leluc}, 
 $R=O_{\mathbb{P}}\left(M^{-1/2-\beta/d}\right)$, the remainder term is negligible compared to $\widehat \Sigma$, which is shown below to be $\sqrt{M}-$asymptotically normal. Let $0<\underline \sigma := \inf_{t\in\mathcal T } \sigma_\eta$ and 
 $\overline \sigma := \sup_{t\in\mathcal T } \sigma_\eta<\infty $. The proof for the asymptotic normality of $\widehat \Sigma$ is  decomposed into several steps. 
 
 \medskip 
 	 
 	\textit{Step 1: Bounds for the moments of $w^{\rm (NN)}_{m}$.}
  	Since $\eta$ and $T$ are independent random variables, we have
 	\begin{equation}\label{eq:var-indiv-noise}
\mathbb{E}\left[\left\{w_m^{\rm (NN)}\right\}^2 \sigma^2_\eta(T_m) \eta_m^2\right] \leq  \overline \sigma^2 \mathbb{E}\left[\left\{w_m^{\rm (NN)}\right\}^2\right]\mathbb{E}\left[\eta_m^2\right] = \overline \sigma^2 \mathbb{E}\left[\left\{w_m^{\rm (NN)}\right\}^2\right].
 	\end{equation}
Recall the notation $\mathbb{E}_M[\cdot] = \mathbb{E}[\cdot \mid M]$. Noting that by construction $\mathbb{E}_M[\widehat d_m] = 1$ and $\mathbb{E}_M[V_m] = M^{-1} $, we obtain 
 	\begin{equation}\label{eq:var-indiv-noise-expanded}
 		\begin{aligned}
 			\mathbb{E}_M\left[\left\{w_m^{\rm (NN)}\right\}^2 \right] &= \frac{1}{M^2}\mathbb{E}_M\left[\left(1 + MV_m - \widehat d_m \right)^2\right] \\
 			&= \frac{1}{M^2} \mathbb{E}_M\left[\left(MV_m -\widehat d_m \right)^2 \right] + \frac{1}{M^2}\\
 			&= \mathbb{E}_M\left[V_m^2 \right] - \frac{2}{M}\mathbb{E}_M\left[V_m \widehat d_m \right] + \frac{1}{M^2}\mathbb{E}_M[\widehat d_m^2] + \frac{1}{M^2}.
 		\end{aligned}
 	\end{equation}
 	On the one hand, by \cite[Theorems 2.1 and 3.1]{devroye2017}, it holds $\lim_{M\rightarrow \infty }M^k\mathbb{E}[V^k_m]=\alpha(d,k)\in (0,\infty) $, for some constant $\alpha(d,k)$ depending on $k$ and the dimension $d$. On the other hand, by \cite[Lemma 1.3]{henze87}, the degree $\widehat d_m$ is bounded for a fixed dimension $d$. From these facts and the Cauchy-Schwarz inequality, we get
 	\begin{equation}\label{ineq_step1}
 M^{-2}\leq 	\mathbb{E}_M\left[\left\{w_m^{\rm (NN)}\right\}^2 \right]  \lesssim M^{-2}.	
 	\end{equation}
 	
\medskip
 	
\textit{Step 2: Conditional Central Limit Theorem with the weights $w^{\rm (NN)}_{m}$.} 	
We will first show that conditionally given the design points $\mathbf{T} =    (T_1, \dots, T_M)$, such that 
\begin{equation}\label{good_design}
W^2_M:=	 \frac{\sum_{m=1}^M \left|w^{\rm (NN)}_{m}\right|^2 }{\max_{1 \leq m \leq M} \left|w^{\rm (NN)}_m \right|^2} \longrightarrow \infty, \quad \text{ as } M\rightarrow \infty,
\end{equation}
the Lindeberg CLT holds for $\widehat \Sigma$ defined as in \eqref{def_Sig}, but with the $w_m^{\rm (NN)}$ instead of the $w_m$. For now let 
$$
s_M^{\rm (NN)} =  \left[\sum_{m=1}^M \left\{w_m^{\rm (NN)}\right\}^2\sigma^2_{\eta}(T_m)\right]^{1/2}.
$$
Moreover, let the notation $\mathbb{E}_{M, \mathbf{T}}[\cdot] = \mathbb{E}[\cdot \mid M, T_1, \dots, T_M]$.
We check Lindeberg's condition.  Let $\epsilon >0$ and let $\mathbbm{1}\{\cdot\}$ denote the indicator function. Since the design points $\mathbf{T}$ and the $\eta_m$, $ 1 \leq m \leq M$ are mutually independent, we have
 	\begin{multline}
\mathbb{E}_{M, \mathbf{T}} \left[\left\{w_m^{\rm (NN)}\right\}^2 \sigma^2_\eta(T_m) \eta_m^2 \mathbbm{1}\big\{|w_m^{\rm (NN)} \sigma_\eta(T_m) \eta_m| > \epsilon s^{\rm (NN)}_{M }\big\}  \right] \\ = \left\{w_m^{\rm (NN)}\right\}^2 \sigma^2_\eta(T_m) \mathbb{E}_{M, \mathbf{T}} \left[ \eta_m^2 \mathbbm{1}\big\{|w_m^{\rm (NN)} \sigma_\eta(T_m) \eta_m| > \epsilon s^{\rm (NN)}_{M }\big\}   \right]  \\
 			\leq \left\{w_m^{\rm (NN)}\right\}^2 \sigma^2_\eta(T_m)  \times \mathbb{E}_{M, \mathbf{T}} \left[ \eta^2 \mathbbm{1}\big\{|  \eta| > \epsilon (\underline \sigma /\overline \sigma) W_M\big\}   \right].
 	\end{multline}
By \eqref{good_design} and the fact that $\eta$ has a finite variance, we get 
$$
\forall \epsilon >0,\qquad \mathbb{E}_{M, \mathbf{T}} \left[ \eta^2 \mathbbm{1}\big\{|  \eta| > \epsilon (\underline \sigma /\overline \sigma) W_M\big\}   \right]\longrightarrow 0, \quad \text{as } M\rightarrow \infty.
$$
The Lindeberg condition for CLT follows, and we get $\{s_M^{\rm (NN)}\}^{-1} \widehat \Sigma ^{{\rm( NN) }} \stackrel{d}{\longrightarrow} \mathcal{N}(0, 1) $, conditionally on the design satisfying \eqref{good_design}, where $\widehat \Sigma ^{{\rm (NN)}} := \sum_{m=1}^M w^{{\rm (NN)}}_m \sigma_\eta(T_m)\eta_m$.
	
	\medskip

\textit{Step 3: Integrating out design points.} Assume for the moment that 
\begin{equation}\label{good_design2}
W^2_M:=	 \frac{\sum_{m=1}^M \left|w^{\rm (NN)}_{m}\right|^2 }{\max_{1 \leq m \leq M} \left|w^{\rm (NN)}_m \right|^2} \longrightarrow \infty, \quad \text{ in probability} .
\end{equation}
 Let 
$$
\Phi_{M, \mathbf{T}} (u; \widehat \Sigma^{\rm (NN)}) = \mathbb E_{M, \mathbf{T}} \left[ \exp\left(\sqrt{-1} \; u \{s_M^{\rm (NN)}\}^{-1}  \widehat \Sigma^{\rm (NN)}\right)\right],\qquad u\in\mathbb R,
$$ 
be the conditional characteristic function of $\widehat \Sigma ^{{\rm (NN)}} / s^{{\rm (NN)}} _{M}$ given the design points. 
By Step 2, we get 
	\begin{equation}\label{cvg1a}
 		\lim_{M\rightarrow \infty }\Phi_{M, \mathbf{T}} (u; \widehat \Sigma^{\rm (NN)}) = \exp(-u^2/2) , \qquad \forall u \in \mathbb R,
 	\end{equation}
 	provided the sequence of design points satisfies  \eqref{good_design}. 
If \eqref{good_design2} holds true, since  the convergence in probability is characterized by the fact that every sub-sequence has a further sub-sequence which convergences almost surely, we deduce that the convergence in \eqref{cvg1a} holds in probability. Next, by the Dominated Convergence Theorem for a sequence of bounded random variables convergent in probability, we get
$$
\mathbb E_M \left[\Phi_{M, \mathbf{T}} (u; \widehat \Sigma^{\rm (NN)}) \right] = \mathbb E_{M} \left[ \exp\left(\sqrt{-1} \; u \{s_M^{\rm (NN)}\}^{-1}  \widehat \Sigma^{\rm (NN)}\right)\right]  \longrightarrow  \exp(-u^2/2) , \qquad \forall u \in \mathbb R,
$$
which means  $\{s_M^{\rm (NN)}\}^{-1} \widehat \Sigma ^{{\rm( NN) }} \stackrel{d}{\longrightarrow} \mathcal{N}(0, 1) $.

\medskip

\textit{Step 4: Checking condition \eqref{good_design2} for $\mathcal T = [0,1]^d$.} It is shown in the Supplementary Material that
\begin{equation}\label{rates_SM1}
\frac{1}M \sum_{m=1}^M \left\{ \left|Mw^{\rm (NN)}_{m}\right|^2 - \mathbb E \left(\left|Mw^{\rm (NN)}_{m}\right|^2 \right) \right\} = O_{\mathbb P} (M^{-1/2}).
\end{equation}
This and \eqref{ineq_step1} imply 
\begin{equation}\label{rates_SM1b}
 M^{-1}\sum_{m=1}^M  \left|Mw^{\rm (NN)}_{m}\right|^2 \geq 1+ O_{\mathbb P}(M^{-1/2}) = O_{\mathbb P}(M^{-1/2}) . 
\end{equation}	
On the other hand, it is shown in the Supplementary Material that
\begin{equation}\label{rates_SM2}
	\max_{1\leq m \leq M}  \left|Mw^{\rm (NN)}_{m}\right|^2 =  O_{\mathbb P}(M^{a}).
\end{equation}
Taking $0<a < 1/2$ in  \eqref{rates_SM2}, the condition \eqref{good_design2} follows. We conjecture that the condition \eqref{good_design2} holds also for the case where $\mathcal T^d$ is the unit sphere, but leave the justification for future work. 
\medskip

\textit{Step 5: Showing that $ s_M^{-1} \widehat \Sigma  - \{s_M^{\rm (NN)}\}^{-1} \widehat \Sigma ^{{\rm( NN) }} = o_{\mathbb P}(1)$.} To complete the proof it remains to show that the difference between the integration rules based on $w_m$ and $w_m^{(NN)}$ is negligible. Let us note that 
$$
M\left[w^{\rm (NN)}_{m} -w_{m}  \right] =  M V_{m} - \widehat{c}_{m} = 
V_m +  \sum_{j:j\neq m} \left\{ V_m - V^{(j)}_{m} \right\}.
$$
Recall that a point can be the nearest neighbor of at most $\mathfrak C$ points, where $\mathfrak C$ is a constant depending only on the domain and the distance $d(\cdot,\cdot)$, see \cite[Lemma 1.3]{henze87}. Then in the sum in the last display, only at most  $\mathfrak C^\prime$ terms are nonzero, where $\mathfrak C^\prime$ is a constant determined by $\mathfrak C$. Since $\lim_{M\rightarrow \infty }M^k\mathbb{E}[V^k_m]=\alpha(d,k)$, for some positive constant $\alpha(d,k)$, \cite[see Theorems 2.1 and 3.1]{devroye2017}, we get
$$
\mathbb E \left[ \left|w^{\rm (NN)}_{m} -w_{m}  \right|^k \right] \lesssim M^{-2k}.
$$
Moreover, we also have
$$
\mathbb E \left[ \left|w^{\rm (NN)}_{m} +w_{m}  \right|^k \right] \lesssim M^{-k}.
$$
As a consequence, it is shown in the Supplementary Material that 
\begin{equation}\label{eq:rate-Sigma-diff}
\mathbb E \left[\left|\widehat \Sigma ^{{\rm( NN) }} - \widehat \Sigma \right|\right] \lesssim M^{-3/2}.
\end{equation}
Moreover, we have 
\begin{equation}\label{eq:rate-sM-diff}
	\mathbb E \left\{ \left|s_M ^2 - \left\{s_M^{\rm (NN)} \right\}^2\right|\right\}\leq \overline \sigma^2 \sum_{m=1}^M \mathbb E \left[ \left|w^{\rm (NN)}_{m} -w_{m}  \right|
	\left|w^{\rm (NN)}_{m} +w_{m}  \right| \right] \lesssim M^{-2} \ll \{s_M^{\rm (NN)} \}^2\asymp M^{-1}.
\end{equation}
(Here, $\asymp$ means left side bounded above and below by constants times the
right side.) Gathering facts from \eqref{eq:rate-Sigma-diff} and \eqref{eq:rate-sM-diff}, we deduce $ s_M^{-1} \widehat \Sigma  - \{s_M^{\rm (NN)}\}^{-1} \widehat \Sigma ^{{\rm( NN) }} = o_{\mathbb P}(1)$.  We conjecture that this holds also for the case where $\mathcal T$ is the unit sphere, but leave the justification for future work. Finally, the asymptotic approximation of $s_M^{2}$ in the case $\mathcal T = [0,1]$ is proved in the Supplementary Material. The proof is now complete. 
\end{proof}

\bibliographystyle{apalike}
\bibliography{arxiv_refs_PW}

\end{document}